%% file: smefit-code.tex
\documentclass[11pt,a4paper]{article}
\usepackage[colorlinks=true, linkcolor=black!50!blue, urlcolor=blue, citecolor=blue, anchorcolor=blue]{hyperref}
\usepackage[font=small,labelfont=bf,margin=0mm,labelsep=period,tableposition=top]{caption}
\usepackage[a4paper,top=3cm,bottom=2.5cm,left=2.5cm,right=2.5cm,bindingoffset=0mm]{geometry}

\usepackage{graphicx,placeins}
\usepackage{float}
\usepackage{afterpage}
\usepackage{epsfig,cite}
\usepackage{amssymb}
\usepackage{amsmath}
\usepackage{dsfont}
\usepackage{multirow}
\usepackage{url}
\usepackage{xcolor,colortbl}
\usepackage{float}
\usepackage{afterpage}
\usepackage{url}
\usepackage{hyperref}
\usepackage{booktabs}
\usepackage{mathrsfs}

\usepackage{tikz}
\usepackage{tikz-3dplot}
\usepackage[compat=1.0.0]{tikz-feynman}

\usepackage{enumitem}
\usepackage{hyperref}
\usepackage{cite}

\usepackage{pifont}

\usetikzlibrary{shapes, arrows}
\usetikzlibrary{decorations.pathreplacing}
\usetikzlibrary{positioning, calc}
\tikzstyle{fitted} = [rectangle, minimum width=5cm, minimum height=1cm, text centered, draw=black, fill=red!30]
\tikzstyle{operations} = [rectangle, rounded corners, minimum width=2cm,text centered, draw=black, fill=red!30]
\tikzstyle{roundtext} = [rectangle, rounded corners, minimum width=2cm, minimum height=0.8cm, text centered, draw=black, fill=red!30]
\tikzstyle{n3py} = [rectangle, rounded corners, minimum width=3cm, minimum height=1cm, text centered, draw=black, fill=green!30]
\tikzstyle{myarrow} = [thick,->,>=stealth]
\tikzstyle{line} =[draw, -latex']
\tikzstyle{decision} = [diamond, draw, fill=red!20, text width=7.5em, text centered,  inner sep=0pt, minimum height=2em, aspect=4]
\tikzstyle{cloud} = [draw, ellipse,fill=green!20, minimum height=2em]
\tikzstyle{inout} = [rectangle, draw, fill=green!20, text width=9.5em, text centered, rounded corners, minimum height=2em, minimum width=10em]
\tikzstyle{block}=[rectangle, draw, fill=blue!20, text width=9.5em, 
                   text centered, rounded corners, minimum height=2em, 
                   minimum width=10em]

\definecolor{darkgreen}{rgb}{0.0, 0.5, 0.13}

\newcommand{\be}{\begin{equation}}
\newcommand{\ee}{\end{equation}}
\newcommand{\bea}{\begin{eqnarray}}
\newcommand{\eea}{\end{eqnarray}}
\newcommand{\bi}{\begin{itemize}}
\newcommand{\ei}{\end{itemize}}
\newcommand{\ben}{\begin{enumerate}}
\newcommand{\een}{\end{enumerate}}

\newcommand{\lc}{\left[}
\newcommand{\rc}{\right]}
\newcommand{\lp}{\left(}
\newcommand{\rp}{\right)}

\newcommand{\smefit}{{\sc\small SMEFiT }}

\def\gsim{\mathrel{\rlap{\lower4pt\hbox{\hskip1pt$\sim$}}
    \raise1pt\hbox{$>$}}}         
\def\lsim{\mathrel{\rlap{\lower4pt\hbox{\hskip1pt$\sim$}}
    \raise1pt\hbox{$<$}}}         

\newcommand{\cov}{\mathrm{cov}}

\newcommand{\draft}[1]{}

\def\beq{\begin{equation}}
\def\eeq{\end{equation}}

\def\lapprox{\lower .7ex\hbox{$\;\stackrel{\textstyle <}{\sim}\;$}}
\def\gapprox{\lower .7ex\hbox{$\;\stackrel{\textstyle >}{\sim}\;$}}

\usepackage{tabularx}
\newcolumntype{C}[1]{>{\centering\arraybackslash}p{#1}}

\usepackage{amsmath}
\usepackage{amsfonts}
\usepackage{amssymb}
\usepackage{dsfont}
\usepackage{pifont}
\usepackage{booktabs}
\usepackage{graphicx}
\usepackage{epstopdf}
\usepackage{epsfig}
\usepackage{framed}
\usepackage{makeidx}
\usepackage{siunitx}
\usepackage[capitalise]{cleveref}
\usepackage{hyperref}
\usepackage{placeins}
\usepackage[font=small,labelfont=bf]{caption}

\begin{document}
\newgeometry{top=1.5cm,bottom=1.5cm,left=1.5cm,right=1.5cm,bindingoffset=0mm}

\vspace{-2.5cm}
\begin{flushleft}
\begin{figure}[h]
  \includegraphics[width=0.32\textwidth]{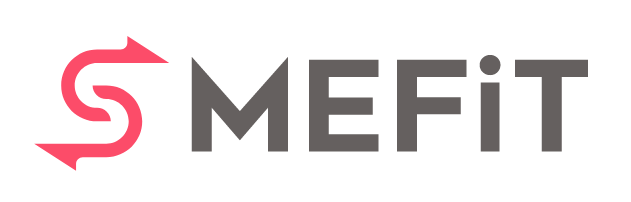}
\end{figure}
\end{flushleft}
\vspace{-3cm}
\begin{flushright}
Nikhef-2022-023 \\
\end{flushright}
\vspace{0.7cm}
\begin{center}
  {\LARGE \bf\boldmath {\sc \LARGE SMEFiT}: a flexible toolbox for  global interpretations\\[0.3cm]
    of particle physics data with effective field theories }\vspace{1.4cm}

Tommaso Giani, Giacomo Magni, and Juan Rojo

\vspace{0.7cm}
{\it \small
Department of Physics and Astronomy, Vrije Universiteit Amsterdam,\\ NL-1081 HV Amsterdam, The Netherlands\\[0.1cm]
Nikhef Theory Group, Science Park 105, 1098 XG Amsterdam, The Netherlands\\[0.1cm]
}

\vspace{0.7cm}

{\bf \large Abstract}

\end{center}

The Standard Model Effective Field Theory (SMEFT) provides a robust
framework to interpret experimental measurements in the context of new physics scenarios
while minimising assumptions on the nature of the underlying UV-complete theory.
We present the {\sc\small Python} open source {\sc\small SMEFiT}  framework,
designed to carry out  parameter inference
in the SMEFT within a global analysis of particle physics data.
{\sc\small SMEFiT} is suitable for inference problems involving
a large number of EFT degrees
of freedom, without restrictions on their functional dependence in the fitted observables,
can include UV-inspired restrictions in the parameter space, and
implements arbitrary rotations between operator bases.
Posterior distributions
are determined from two complementary approaches, Nested Sampling and Monte Carlo optimisation.
{\sc\small SMEFiT} is released together with documentation,
tutorials, and post-analysis reporting tools,
and can be used to carry out state-of-the-art
EFT fits of Higgs, top quark, and electroweak production data.
To illustrate its functionalities,  we reproduce the results
of the recent ATLAS EFT interpretation
of Higgs and electroweak data from Run II and demonstrate how
equivalent results are obtained in two different operator bases.

\clearpage
\tableofcontents

\input{sec-introduction}
\input{sec-framework}
\input{sec-atlaseft}

\input{sec-summary}

\FloatBarrier

\bibliography{smefit-code}

\end{document}

%% file: sec-introduction.tex
\section{Introduction}
\label{sec:intro}

Global interpretations of particle physics observables in the framework
of the Standard Model Effective Field Theory
(SMEFT)~\cite{Weinberg:1979sa,Buchmuller:1985jz,Grzadkowski:2010es}, see~\cite{Hartland:2019bjb,Brivio:2019ius,Biekotter:2018rhp,Ellis:2018gqa,Almeida:2018cld, Aebischer:2018iyb,Ellis:2020unq,Bissmann:2020mfi,Bruggisser:2021duo,Bruggisser:2022rhb}
for recent analyses,
require the inference of several tens or eventually hundreds
of independent Wilson coefficients from experimental data.
For instance, the {\sc\small SMEFiT} analysis~\cite{Ethier:2021bye} of Higgs, top, and diboson data
from the LHC constrains  36 independent directions in the SMEFT parameter space,
with 14 more coefficients fixed indirectly from electroweak precision observables.
Exploring efficiently such a broad parameter space is only possible
by means of the combination of a large number of measurements
from different processes with state-of-the-art theoretical calculations.
Realizing the ultimate goal of the SMEFT paradigm, a theory-assisted combination
of measurements from the high-energy frontier down to the electroweak scale,
flavour physics, and low-energy observables, demands a flexible and robust EFT fitting methodology
amenable for inference problems involving up to hundreds of coefficients from
a wide range of different physical observables,
each of them characterised by their own statistical model.

In addition to being able to constrain such large parameter spaces,
a SMEFT fitting framework suitable to the aforementioned goal
ought to satisfy several other requirements.
These include, but are not restricted to: carrying out fits both at
the linear and at the quadratic
level in the EFT expansion;
accounting for all relevant sources of methodological, theoretical, and experimental
uncertainties;
exhibiting a  modular structure enabling the seamlessly incorporation of
new processes or improved theory calculations;
accepting general likelihood functions
beyond  
the multi-Gaussian approximation like those associated to unbinned
or Poission-distributed observables;
providing  statistical and visualization diagnosis tools to assist
the  interpretation of the results, from PCA
and  Fisher Information to basis rotation and reduction
algorithms; and
implementing theoretical constraints on the parameter space, such
as those associated to the  matching
to UV-complete scenarios.
Furthermore, the availability of such a fitting tool as open source would
facilitate its adoption by interested parties
both from the theory and the experimental communities.

Several fitting frameworks have been developed
and deployed in the context of SMEFT interpretations of particle physics data,
such as {\sc\small SMEFiT}~\cite{Hartland:2019bjb}, {\sc\small FitMaker}~\cite{Ellis:2020unq},
{\sc\small HepFit}~\cite{DeBlas:2019ehy}, {\sc\small EFTfitter}~\cite{Castro:2016jjv},
and {\sc\small Sfitter}~\cite{Brivio:2019ius}
among others.
{\sc\small SMEFiT} was originally developed in the context of top quark studies~\cite{Hartland:2019bjb},
and then extended to Higgs and diboson measurements in~\cite{Ethier:2021bye}
and to vector boson scattering in~\cite{Ethier:2021ydt}.
Whenever possible, these analyses strived to account for
NLO QCD calculations in the EFT cross-sections such
as those provided by the SMEFT@NLO package~\cite{Degrande:2020evl}.
Additional studies based on {\sc\small SMEFiT} are the implementation of the Bayesian
reweighting method~\cite{vanBeek:2019evb} and the LHC EFT WG report on experimental
observables~\cite{Castro:2022zpq}.
Advantages of {\sc\small SMEFiT} as compared to related frameworks
include  independent and complementary statistical
methods to carry out parameter inference
(Monte Carlo optimisation and Nested Sampling), the lack of restrictions
on the allowed functional dependence in $c_i/\Lambda^2$  for the fitted observables,
and a competitive scaling of the running time with the number of fitted parameters.

The goal of this paper is to present and describe the release of
{\sc\small SMEFiT} as a {\sc\small Python} open source fitting framework, together with the corresponding
datasets and theory calculations required to reproduce published analyses.
{\sc\small SMEFiT} is made available via its public GitHub repository
\begin{center}
\url{https://github.com/LHCfitNikhef/smefit_release}
\end{center}  
together with the documentation and user-friendly tutorials provided in
\begin{center}
\url{https://lhcfitnikhef.github.io/smefit_release}
\end{center}
which also includes a catalog of SMEFT analyses corresponding to different choices
of the input dataset, theoretical settings, and statistical methodology.
This online documentation is the main resource to consult in order
to use {\sc\small SMEFiT}, either to reproduce existing analyses or to extend them
to new processes and observables, and hence here we restrict ourselves to highlighting
 key representative
results.
Technical aspects of the framework, such as the format
in which the data and the theory calculations are to be provided, are already
described in the online documentation and therefore are not covered in this paper.

Here first we reproduce the results of the global SMEFT analysis of~\cite{Ethier:2021bye}.
In doing so,
we fix a number of small issues that were identified during the code rewriting process.
Then we illustrate the capabilities of {\sc\small SMEFiT} by independently reproducing the ATLAS
EFT interpretation of Higgs and electroweak data from the LHC, together with LEP measurements,
presented in~\cite{ATL-PHYS-PUB-2022-037}.
We demonstrate how when using the same inputs in terms of experimental data and
EFT parametrisation one obtains the same bounds in the Wilson coefficients.
Furthermore, we show how results are independent of the choice of fitting basis:
equivalent results are obtained when
 using either the Warsaw basis or the rotated basis chosen in~\cite{ATL-PHYS-PUB-2022-037}
to restrict the parameter space to directions with large variability.

This benchmark exercise displays the capabilities of {\sc\small SMEFiT} to contribute to
the ongoing and future generation of SMEFT studies, where the careful comparison
between the outcomes from different groups is instrumental to cross-check
independent determinations.
This program, partly carried out in the context of the LHC EFT WG
activities, aims to bring the robustness of SMEFT analyses on par to that
of SM calculations, simulations, and benchmark comparisons.

The outline of this paper is as follows.
Sect.~\ref{sec:framework} describes the  {\sc\small SMEFiT} framework
and validates the code rewriting by comparing its outcome
with that of the global SMEFT analysis of~\cite{Ethier:2021bye}.
Sect.~\ref{sec:atlaseft} illustrates the possible applications
of {\sc\small SMEFiT} by reproducing
the results of the ATLAS EFT fit of LHC and LEP data from~\cite{ATL-PHYS-PUB-2022-037}
and demonstrating the fitting basis independence of the results.
We conclude and outline possible future developments in Sect.~\ref{sec:summary}.

%% file: sec-framework.tex
\section{The \smefit framework}
\label{sec:framework}
\label{sec:settings}

In this section we provide a concise overview of the main features
and functionalities of the \smefit framework, pointing the reader
to the original publications~\cite{Hartland:2019bjb,Ethier:2021bye,Ethier:2021ydt,vanBeek:2019evb}
as well as to the code online documentation for more details.

\paragraph{Installation.}
The \smefit code can be installed by using the {\sc\small conda} interface.
An installation script is provided, allowing the user to create a {\sc\small conda} environment 
compatible with the one which has been automatically tested,  and where the \smefit package 
can be installed and executed.
{\sc \small conda} lock files ensure that results are always produced using the 
correct version of the code dependencies, regardless of the machine where the environment is created, 
hence ensuring complete reproducibility of the results.
In this same environment the code can be easily edited,
allowing the users to contribute to the development of the open-source framework. 

\paragraph{EFT cross-section parametrisation.}
In the presence of $n_{\rm op}$ dimension-six SMEFT operators, a general SM cross-section 
$\sigma_{\rm SM}$ will be modified as follows
\begin{align}
    \label{eq:eft_param}
   \sigma_{\rm eft}\lp {\boldsymbol c}/\Lambda^2\rp =  \sigma_{\rm SM} +  \sum_{i=1}^{n_{\rm op}} \tilde{\sigma}_{{\rm eft},i} \frac{c_i}{\Lambda^2} 
    + \sum_{i,j=1}^{n_{\rm op}} \tilde{\sigma}_{{\rm eft},ij} \frac{c_ic_j}{\Lambda^4} \, ,
\end{align}
with $\tilde{\sigma}_{{\rm eft},i}/\Lambda^2$ and $\tilde{\sigma}_{{\rm eft},ij}/ \Lambda^4$ 
indicating respectively the contributions to the cross-section
arising from the interference with the SM amplitudes and from the square of the EFT ones,
once the Wilson coefficients $c_i$ are factored out.
These cross-sections hence depend on $n_{\rm op}$ Wilson coefficients and on the cutoff scale
$\Lambda$, with only the ratios ${\boldsymbol c}/\Lambda^2$ being accessible in a model-independent analysis.
Eq.~(\ref{eq:eft_param}) can be generalised when other types of SMEFT operators, e.g. dimension-eight operators,
are considered in the interpretation of the observable.

The terms $\sigma_{\rm SM} $, $\tilde{\sigma}_{{\rm eft},i}$ and $\tilde{\sigma}_{{\rm eft},ij}$
in Eq.~(\ref{eq:eft_param}),
are inputs to \smefit and are provided by means of external calculations.
In this respect, the fitting code is agnostic in the calculational settings
used to produce them, provided they comply with the required format of the theory tables described
below.

The user can also choose to adopt an alternative form for the theory predictions
\begin{align}
    \label{eq:eft_param_2}
   \sigma_{\rm eft}\lp {\boldsymbol c}/\Lambda^2\rp =  \sigma_{\rm SM}\lp 1 +  \sum_{i=1}^{n_{\rm op}} \kappa_{{\rm eft},i} \frac{c_i}{\Lambda^2} 
    + \sum_{i,j=1}^{n_{\rm op}} \kappa_{{\rm eft},ij} \frac{c_ic_j}{\Lambda^4}\rp \, ,
\end{align}
with now the EFT contributions entering as $K$-factors multiplying the SM prediction.
The multiplicative variant in Eq.~(\ref{eq:eft_param_2}) is equivalent to Eq.~(\ref{eq:eft_param}) 
only in cases where higher order QCD and electroweak corrections
coincide in the SM and in the SMEFT.
Eq.~(\ref{eq:eft_param_2}) benefits from reduced
theory uncertainties on the EFT contribution,
such as the one due to PDFs,
which partially cancel when taking the $K$-factor ratios.

\paragraph{Experimental data and theory predictions.}
The theoretical predictions and experimental data for the processes
entering an EFT intepretation are considered as external, 
user-provided inputs to {\sc\small SMEFiT}.
As such, they are stored in the following separate GitHub repository
\begin{center}
  \url{https://github.com/LHCfitNikhef/smefit_database}
\end{center}
since in this way one separates code developments from changes in the external data and theory inputs.
This repository should be cloned separately and then the local path specified in the runcard.
Detailed instructions are given in the online documentation.

\begin{figure}[t]
    \center
    \includegraphics[scale=0.25]{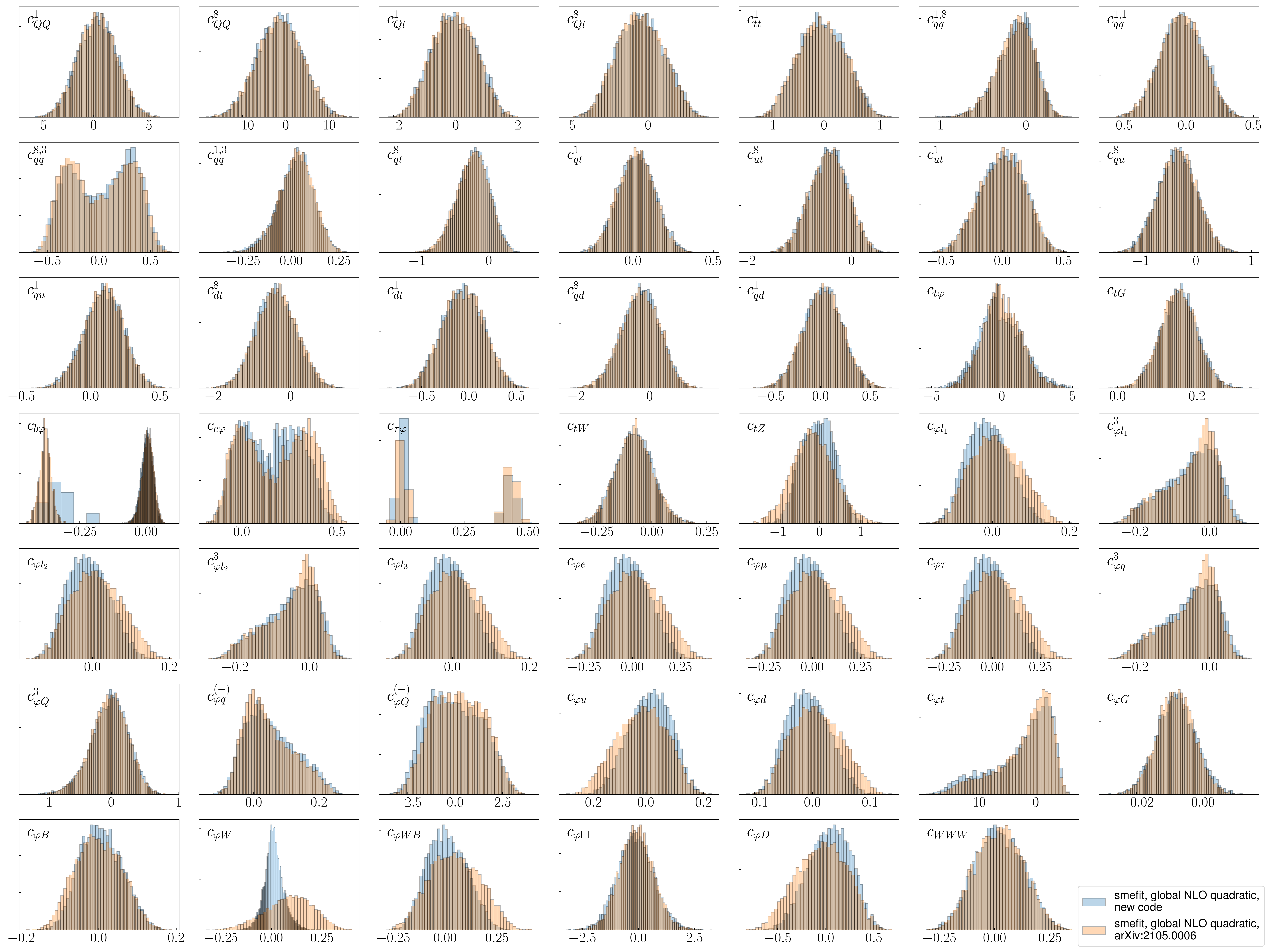}
    \caption{\small The posterior distributions in the SMEFT parameter space
      from the \smefit global analysis of~\cite{Ethier:2021bye} compared to those
      obtained from the new code for the same theory settings and datasets,
      using in  both cases NS and EFT cross-sections
      accounting for NLO QCD corrections and quadratic $\mathcal{O}\lp \Lambda^{-4}\rp$ effects.
      The shown posteriors assume $\Lambda = 1\, \text{TeV}$, 
      and can be appropriately rescaled for other values of $\Lambda$.
     }
    \label{fig:new_vs_old}
\end{figure}

Currently, this database repository includes the tables for experimental data and
theory predictions required to reproduce the global SMEFT analysis of~\cite{Ethier:2021bye},
see also Fig.~\ref{fig:new_vs_old},
as well as the ATLAS EFT interpretation of~\cite{ATL-PHYS-PUB-2022-037},
to be discussed in Sect.~\ref{sec:atlaseft}.
This database will be kept updated as additional processes
and improved theory calculations entering the \smefit
global analyses are included.
Users of the code can take the existing datasets as templates for the implementation
of new processes for their own  EFT fits.

Theory predictions are stored in {\tt JSON} format files composed
by a dictionary that contains, for each dataset, the central SM predictions,
the LO and NLO linear and quadratic EFT cross-sections, and the theory
covariance matrix.
For the experimental data instead we adopt a {\tt YAML} format
which contains the number of data points, central values, statistical
errors, correlated systematic errors, and the type of systematic error (additive
of multiplicative), from which the covariance matrix of the measurements can be
constructed.
Alternatively, for datasets in which the breakdown of systematic errors
is not provided, the user has to decompose the covariance matrix into a set of correlated systematic
errors.
Details regarding the format of data, uncertainties, and theory predictions are provided in 
the corresponding section of the code documentation
\begin{center}
  \url{https://lhcfitnikhef.github.io/smefit_release/data_theory/data.html}
\end{center}

The \smefit runcard which steers the code should list the experimental inputs that enter
the fit and the corresponding theory calculations, including the path to the folders
where these inputs are stored.
In the code repository one can find examples of runcards that can be used
to reproduce the two EFT interpretations mentioned above, together with the corresponding
post-fit analysis reports.

\paragraph{Likelihood function.}
The goal of \smefit is to determine confidence level intervals in the space of the Wilson coefficients
given an input dataset $\mathcal{D}$ and the corresponding theory predictions
$\mathcal{T}(\boldsymbol{c})$, with the latter given by Eq.~(\ref{eq:eft_param})
or generalizations thereof.
The agreement between the dataset $\mathcal{D}$ and a theory hypothesis $\mathcal{T}(\boldsymbol{c})$
is quantified by the likelihood function $\mathcal{L}( \boldsymbol{c})$.
The wide majority of measurements used in EFT interpretations are presented
as multi-Gaussian distributions, for which the likelihood is given by
\begin{equation}
  \label{eq:binned_gaussian_likelihood}
  \mathcal{L}(\boldsymbol{c}) \propto \prod_{m,n=1}^{n_{\rm dat}} \exp\left[-\lp
  \mathcal{T}_m(\boldsymbol{c}) - \mathcal{D}_m\rp \lp {\rm cov}^{-1} \rp_{mn}
    \lp
    \mathcal{T}_n(\boldsymbol{c}) - \mathcal{D}_n\rp\right] \, .
\end{equation}
In such cases, the log-likelihood function $(-\log \mathcal{L})$ becomes either 
a quadratic or a quartic function of the Wilson coefficients ${\boldsymbol c}$, 
depending on whether the quadratic terms in the EFT parametrization of Eq.~(\ref{eq:eft_param}) 
are retained, according to the user specifications in the runcard.

Despite Eq.~\eqref{eq:binned_gaussian_likelihood} being the only 
functional form for the likelihood which is currently implemented in {\sc\small SMEFiT}, 
the modular structure of the code can be easily extended to accomodate alternative likelihood functions.
For instance, for unbinned observables~\cite{GomezAmbrosio:2022mpm}
the likelihood would receive a contribution of the form
\begin{equation}
  \mathcal{L}( \boldsymbol{c}) = \frac{\nu_{\rm tot}(\boldsymbol{c})^{N_{\rm ev}}}{N_{\rm ev}!}e^{-\nu_{\rm tot}(\boldsymbol{c})}\prod_{i=1}^{N_{\rm ev}} \frac{1}{\sigma_{\rm fid}(\boldsymbol{c}) }\frac{d\sigma(\boldsymbol{x}, \boldsymbol{c})}{d\boldsymbol{x}}
  \, ,
\label{eq:likelihood_extended}
\end{equation}
with $N_{\rm ev}$ being the number of events, $\nu_{\rm tot}$ the expected event count from theory,
and the event probability is determined by the cross-section differential in the event kinematics
$\boldsymbol{x}$.
Such unbinned likelihood could be implemented in {\sc\small SMEFiT},
allowing for a general EFT interpretation involving a combination of measurements each of
which described by a different statistical model.

\paragraph{Nested Sampling.}
Once a statistical model $\mathcal{L}( \boldsymbol{c})$ for the input dataset $\mathcal{D}$
is defined in terms of the theory predictions $\mathcal{T}(\boldsymbol{c})$,
 \smefit proceeds to determine the most likely values of the Wilson coefficients
$\boldsymbol{c}$ and the corresponding uncertainties by means of two different and complementary
strategies.
The first one is based on Nested Sampling (NS) via the MultiNest
library~\cite{Feroz:2013hea,Feroz:2007kg}.
The installation of the latter is automatically carried out by the \smefit installation script.

The starting point is Bayes' theorem
relating the posterior probability distribution of
the parameters $\boldsymbol{c}$ given the observed data
and the theory hypothesis, $P\lp\boldsymbol{c}| \mathcal{D},\mathcal{T} \rp $,
to the likelihood function (conditional probability)
and the prior distribution $ \pi$,
\be
\label{eq:bayestheorem}
P\lp\boldsymbol{c}| \mathcal{D},\mathcal{T} \rp= \frac{P\lp\mathcal{D}|\mathcal{T},\boldsymbol{c}
  \rp P\lp \boldsymbol{c}|\mathcal{T}  \rp
}{P(\mathcal{D}|\mathcal{T})} = \frac{\mathcal{L}\lp\boldsymbol{c} \rp
\pi \lp  \boldsymbol{c} \rp}{\mathcal{Z}} \, ,
\ee
with the  Bayesian evidence $\mathcal{Z}$  ensuring the normalisation of the posterior
distribution,
\be
\label{eq:bayesianevidence}
\mathcal{Z} = \int \mathcal{L}\lp  \boldsymbol{c} \rp
\pi \lp  \boldsymbol{c} \rp d \boldsymbol{c} \, .
\ee
By means of Bayesian inference, NS maps the $n_{\rm op}$-dimensional integral over the prior density
into 
\be
\label{eq:NS1}
X(\lambda) = \int_{\{ \boldsymbol{c} : \mathcal{L}\lp\boldsymbol{c} \rp > \lambda \}}
\pi(\boldsymbol{c} ) d\boldsymbol{c} \,,
\ee
a  one-dimensional function corresponding to the
volume of the prior density $\pi(\boldsymbol{c} )d\boldsymbol{c}$ associated
to likelihood values  greater than  $\lambda$.
Eq.~(\ref{eq:NS1}) defines a transformation between the prior
and posterior distributions sorted by the likelihood 
of each point in the EFT parameter space, is evaluated
numerically, and results in $n_{\rm spl}$ samples $\{ \boldsymbol{c}^{(k)} \}$ providing
a representation of the posterior probability distribution
from which one can evaluate confidence level intervals
and related statistical estimators.
The default \smefit analyses assumes  a flat prior volume $\pi(\boldsymbol{c})$,
although implementing alternative functional forms for the prior volume is an option
available to the user.

One benefit of sampling methods such as NS is that they bypass limitations
of numerical optimisation techniques such as local minima preventing reaching the absolute minimum,
with a drawback being that
the computational resources required in NS grow exponentially with the dimensionality $n_{\rm op}$.

\paragraph{MCfit optimisation.}
The second strategy available in {\sc\small SMEFiT}, denoted by MCfit,
is based on the Monte Carlo replica method used e.g.
by the NNPDF analyses of parton distributions~\cite{DelDebbio:2004xtd,Ball:2008by}.
$N_{\rm rep}$ Gaussian replicas $\mathcal{D}_{n}^{(k)}$ of the experimental data $\mathcal{D}_{n}$,
with $n=1,\ldots,n_{\rm dat}$,
are generated
according to the covariance matrix of Eq.~\eqref{eq:binned_gaussian_likelihood}.
Subsequently, the best-fit coefficients ${\boldsymbol c}^{(k)}$ for each data replica
$\mathcal{D}_n^{(k)}$
are determined from the numerical minimisation of the log-likelihood function. 

Several minimisers are available for this purpose in {\sc\small SMEFiT}: the evolutionary
CMA-ES algorithm~\cite{cmaes} used in the fragmentation function fits of~\cite{Bertone:2017tyb};
and two build-in minimizers provided by {\sc \small scipy}~\cite{Byrd1999AnIP,1997PhLA..233..216X}.
For each of them, the user can specify different settings controlling the efficiency and accuracy 
of the minimisation. 
Additional algorithms can be added by the user.
We note that as opposed to the PDF fit case no cross-validation is required
here, since overlearning is not possible for a discrete parameter space, where the best-fit value
coincides with the absolute maximum of the likelihood.

The final result of MCfit is a sample of $N_{\rm rep}$ replicas $\{ \boldsymbol{c}^{(k)} \}$  that
provides a representation of the probability density in the space of SMEFT coefficients,
and that can be processed in the same manner as its NS counterpart.
While the posteriors obtained with MCfit and NS should be equivalent,
in practice small residual differences can appear and traced back to 
numerical inefficiencies of the minimiser.
In this respect, we recommend that in \smefit the NS method is adopted as baseline,
with MCfit as an independent cross-check.
As compared to NS, the computational performance of MCfit scales better
with $n_{\rm op}$ with the duration of single-replica fits being the limiting factor.

\paragraph{Theoretical uncertainties.}
The covariance matrix that enters the Gaussian likelihood in Eq.~(\ref{eq:binned_gaussian_likelihood})
contains in general contributions of both experimental and theoretical origin.
Assuming that these two sources of uncertainty are uncorrelated and that the latter can be approximated
by a multi-Gaussian distribution~\cite{AbdulKhalek:2019bux,AbdulKhalek:2019ihb},
the covariance matrix used in \smefit is defined by
\be
\label{eq:theory_covmat}
\cov_{nm} = \cov^{(\rm exp)}_{nm} + \cov^{(\rm th)}_{nm} \, , \qquad n,m=1,\ldots,n_{\rm dat} \,,
\ee
namely as the sum of the experimental and the theoretical covariance matrices.
The latter should contain in principle all relevant sources of theory error such as
PDF, missing higher orders (MHO), and
MC integration uncertainties, with MHOU being treated according to the
formalism developed in~\cite{AbdulKhalek:2019bux,AbdulKhalek:2019ihb}.
In practice, theory errors should be specified in the theory tables for each measurement.
Note that the code is agnostic with respect to the source 
of theory errors provided by the user, and in particular can account for the 
PDF and MHO uncertainties associated to the linear and quadratic EFT predictions 
whenever the theory covariance matrix provided in the \smefit theory tables takes these into account.
We note that in the current implementation, correlations between theory uncertainties
corresponding to different datasets are neglected.

\paragraph{Constrained fits.}
Within a SMEFT interpretation of experimental data it is often necessary to impose relations between
some of the fitted Wilson coefficients, rather than keeping all of them as free parameters.
Such constraints in the SMEFT parameter space arise for instance as a consequence of the matching
to specific UV-complete models, but also from
the approximate implementation of electroweak precision observables via
a restriction in the parameter space used in~\cite{Ethier:2021bye}, as well
as from  simplified
EFT interpretations with more restrictive flavour assumptions.
One example of the latter, considered in~\cite{Ethier:2021bye} and proposed by the
LHC Top working group in~\cite{AguilarSaavedra:2018nen}, is that of a top-philic scenario
with new physics coupling preferentially to the top quark.
This scenario is based
on the  assumption that new physics couples dominantly to the left-handed doublets 
and right-handed up-type quark singlet of the third generation as well as to gauge bosons, 
and as compared to  the baseline settings in~\cite{Ethier:2021bye} 
this assumption introduces additional restrictions in the EFT parameter space.

In general, linear constraints can be implemented via the \smefit runcard and lead to a speedup
of the fitting procedure.
The implementation of the same type of constraints a posteriori by means
of the Bayesian reweighting method~\cite{Ball:2011gg} was demonstrated in~\cite{vanBeek:2019evb}, showing that
it leads to a large efficiency loss and hence is only reliable for moderate restrictions in the parameter space.

In several scenarios the matching procedure between the SMEFT
and UV-complete models results in non-linear relations between
the Wilson coefficients.
The automated implementation of such non-linear constraints in \smefit is work in progress and requires non-trivial
modifications of the fit procedure.
An upcoming publication focused on matching to UV-complete models will discuss this problem
in more detail.

\paragraph{Basis selection and rotation.}
The baseline choice for the theory tables containing the linear
and quadratic EFT predictions in \smefit is that these are provided by the user
in the Warsaw basis.
In general, it might be more convenient to carry out the fit in a different basis, for instance
one closer to the actual constraints imposed by the experimental data considered.
The user can thus indicate how the chosen fitting basis is related to the Warsaw operators by means
of a rotation matrix 
\begin{align}
  \label{eq:PCA-rotatedbasis-sec2}
  \mathcal{O}^{\text{(F)}}_{i} = \sum_{j=1}^{n_{\rm op}} R^{\rm{(W\to F)}}_{ij} \mathcal{O}^{\text{(W)}}_{j}\,,\quad  i =1,\ldots,n_{\rm op} \,,  \qquad R^{\rm{(F\to W)}}= \lp R^{\rm{(F\to W)}}\rp^{-1} \, ,
\end{align}
with $\mathcal{O}_i^{(W)}$ and $\mathcal{O}_j^{(F)}$ indicating the operators in
the Warsaw and fitting bases respectively.
Note that the number of operators can be different in the two bases considered, or more precisely,
in the fit basis a number of operators can be set to zero, for instance when unconstrained
by the data.

This rotation matrix can also be determined automatically from a principal component analysis (PCA)
of the Fisher information matrix (defined below) which determines the directions with the highest variability.
EFT directions with the lowest variability can be set to zero as a constrain in order to remove
quasi-flat directions and thus increase the numerical stability of the fits.
Results of an EFT interpretation should of course be basis independent, provided that the two bases are related
by a rotation.
We will exploit these functionalities of the \smefit framework in Sect.~\ref{sec:atlaseft} to reproduce
the ATLAS EFT interpretation of~\cite{ATL-PHYS-PUB-2022-037} in two different bases
and verify that results are identical.

\paragraph{Fisher information.}
A measure of the sensitivity of individual datasets to specific directions
in the EFT parameter space is provided by the Fisher information  matrix $I_{ij}$, defined as
\be
\label{eq:FisherDef}
I_{ij}\lp {\boldsymbol c} \rp = -{\rm E}\lc \frac{\partial^2 \ln \mathcal{L} \lp
{\boldsymbol c} \rp}{\partial c_i \partial c_j} \rc \, , \qquad i,j=1,\ldots,n_{\rm op} \, ,
\ee
where ${\rm E}\lc~\rc$ indicates the expectation value over the Wilson coefficients and
$ \mathcal{L} \lp {\boldsymbol c} \rp$ is the likelihood function.
The covariance matrix of the Wilson coefficients, $C_{ij} \lp {\boldsymbol c} \rp$,
is  bounded by the Fisher information matrix, $C_{ij} \ge \lp I^{-1}\rp_{ij}$,
the so-called Cramer-Rao bound, which illustrates how $I_{ij}$
quantifies the constraining power of the dataset $\mathcal{D}$.

In the specific case of linear EFT calculations and a diagonal
covariance matrix,
the Fisher information matrix Eq.~(\ref{eq:FisherDef}) simplifies to
\be
\label{eq:fisherinformation2}
I_{ij} = \sum_{m=1}^{n_{\text{dat}}}\frac{\sigma_{{\rm eft},i,m}\, \sigma_{{\rm eft},j,m}}{\delta^2_{\text{tot},m}}\, ,
\qquad i,j=1,\ldots,n_{\rm op} \,
\ee
with $\delta_{\text{tot},m}$ being the total uncertainty of the $m$-th data point,
such that $I_{ij}$
 is independent of the fit results and can be evaluated a priori.
Eq.~(\ref{eq:fisherinformation2})  shows that at the linear EFT level the Fisher information
is the average of the EFT corrections to the SM cross-section in the dataset
$\mathcal{D}$ in units of the measurement uncertainty.
We emphasize that in \smefit we always evaluate Eq.~(\ref{eq:FisherDef}) in terms
of the full covariance matrix and Eq.~(\ref{eq:fisherinformation2}) is provided
only for illustration purposes.

\smefit evaluates the Fisher information matrix Eq.~(\ref{eq:FisherDef}) for the datasets
and theory predictions specified in the runcard, and presents the results
graphically to  facilitates the interpretation of the results.
The absolute normalisation of the Fisher matrix is arbitrary, since one can always
rescale operator normalizations.
Hence we normalise it such
that it becomes independent of the choice of overall operator normalisation.
As mentioned above, the user can apply the  PCA to this Fisher information
matrix to determine the directions (principal components) with highest variability,
and eventually use them as fitting basis, rather than the original Warsaw basis,  by
applying a rotation of the form of Eq.~(\ref{eq:PCA-rotatedbasis-sec2}).

\begin{figure}[htbp]
    \center
    \includegraphics[scale=0.53]{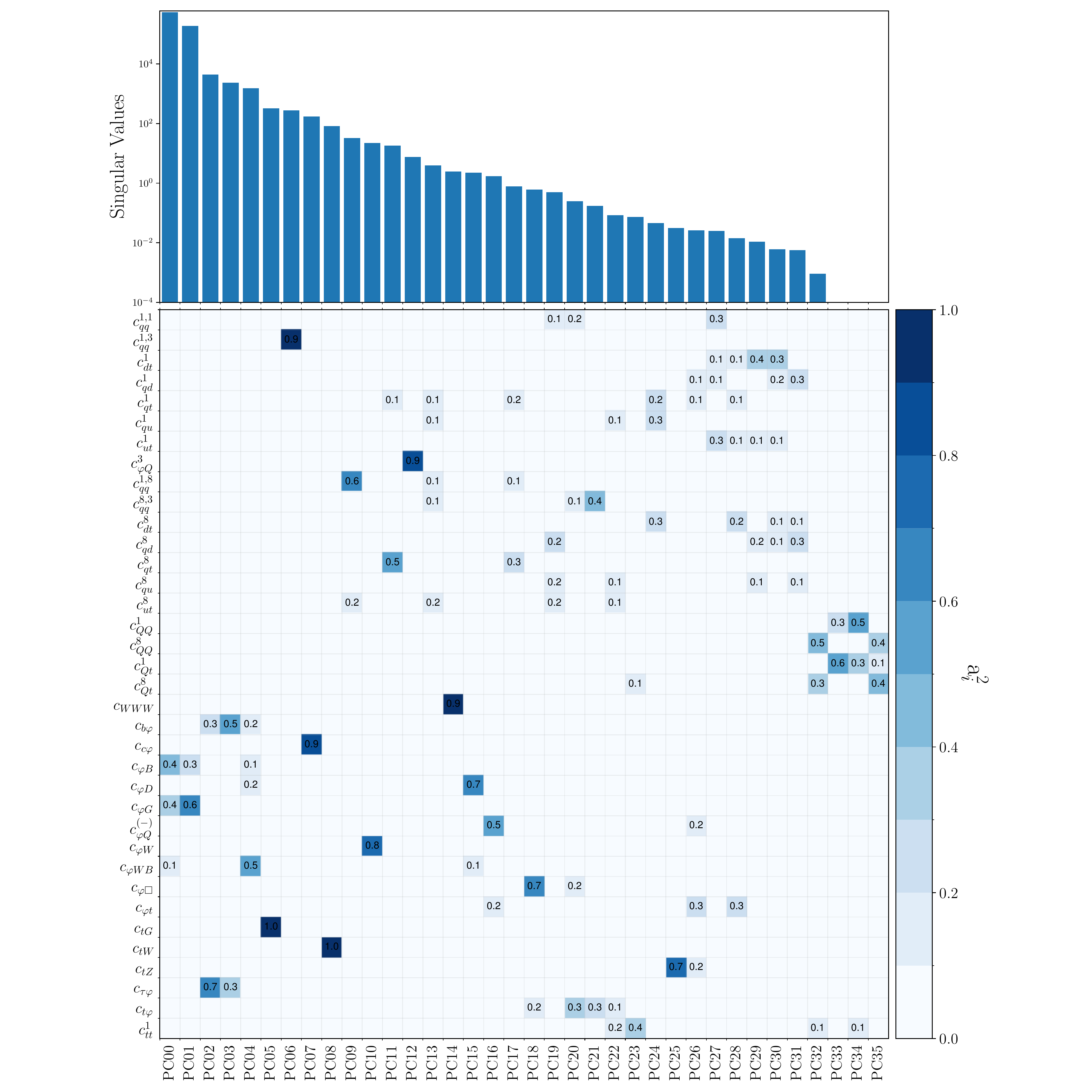}
    \caption{\small Results of the Principal Component Analysis applied to the Fisher information matrix (for linear
      EFT calculations) of the global dataset of~\cite{Ethier:2021bye}.
      For each principal component, we display the coefficients of the linear combination
      of fit basis operators and the corresponding singular value.
      Three flat directions, corresponding to three linear combinations
      of four-heavy-quark operators, have vanishing singular values indicating
      that cannot be constrained from the fit at the linear level.
    }
    \label{fig:pca_heatmap_mult_prescr_lin}
\end{figure}

Fig.~\ref{fig:pca_heatmap_mult_prescr_lin} displays the
 PCA applied to the Fisher information matrix (in the linear
 EFT case) for the global dataset of~\cite{Ethier:2021bye}.
 For each principal component, we display the coefficients of the linear combination
 of fit basis operators and the corresponding singular value.
 Three flat directions, corresponding to three linear combinations
 of four-heavy-quark operators, have vanishing singular values indicating
 that cannot be constrained from the fit.

\paragraph{Fit report and visualization of results.}
The output of a \smefit analysis consists on the posterior probability distributions
associated to the fit coefficients ${\boldsymbol c}$ as well as ancillary statistical estimators such
as the fit quality per dataset.
This output can be processed and visualized by means of a fit report,
which is generated by specifying a separate runcard either for an individual
fit or for pair-wise (or multiple) comparisons between fits.

By modifying this dedicated runcard
the user can specify what to display in the report, with currently available options including
comparisons between SM and best-fit EFT predictions for individual datasets,
posterior distributions with associated correlation and confidence level bar plots,
two-parameter contour plots, the log-likelihood distribution among replicas or samples, Fisher matrix by dataset, and the outcome of the PCA analysis among others.
The online documentation contains the description of the report runcard 
and examples of fits reports obtained with specific runcards,

\begin{center}
  \url{https://lhcfitnikhef.github.io/smefit_release/report/running.html}\,.
\end{center}
The \smefit report is produced both in {\small .pdf} and {\small .html} format 
to facilitate readability and visualization.

\paragraph{Code rewriting and validation.}
As compared to the version of the code used for the global SMEFT analysis of~\cite{Ethier:2021bye},
the \smefit framework has been completely rewritten in preparation for its
public release, streamlining its overall structure and enhancing its modular
character and user-friendly interface.
In this process, both the code itself and the data and theory tables have been
repeatedly cross-checked using the baseline fit of~\cite{Ethier:2021bye} as benchmark.
A number of small issues were identified and corrected in the theory tables, without affecting
any of the main findings of the original study.
We have verified that the only differences between~\cite{Ethier:2021bye} and the results shown 
in Fig.~\ref{fig:new_vs_old} obtained with the new code is related to bug fixes in the theory tables, 
and that if with the new code we use the same theory tables as in~\cite{Ethier:2021bye}, 
identical posterior distributions are obtained both at the linear and the quadratic EFT levels.

To highlight the agreement between the results of~\cite{Ethier:2021bye} and the output of the
new \smefit release code,
Fig.~\ref{fig:new_vs_old} compares the posterior distributions in the SMEFT parameter space
obtained in~\cite{Ethier:2021bye} with those
based on the new version of the code and of the theory and data tables.
Posterior distributions are evaluated with Nested Sampling for the global dataset,
while EFT cross-sections
account for both NLO QCD corrections and for quadratic $\mathcal{O}\lp \Lambda^{-4}\rp$ effects.
The 95\% CL intervals obtained in both cases are very similar, with possibly the exception
of the  $c_{\varphi W}$ bosonic operators where its uncertainty was somewhat overestimated
in the original analysis.
A similar or better level of agreement is found for the linear EFT fits and for the fits
based on LO EFT cross-sections.
Good agreement is also found for related fit estimators, such as the
correlation matrix between Wilson coefficients.

As mentioned above, within the \smefit framework one can choose between two
alternative and complementary strategies to determine CL intervals on the Wilson coefficients
entering the theory calculations, namely Nested Sampling and MCfit.
Each method has its own advantages and disadvantages, for instance MCfit scales better with
the number of fit parameters but may be affected by numerical inefficiencies of the minimisation
procedure, specially for poorly constrained operators.
Within the current framework, we recommend users to adopt NS as the default
strategy and use MCfit as an independent
cross-check.

\begin{figure}[t]
    \center
    \includegraphics[width=0.49\textwidth]{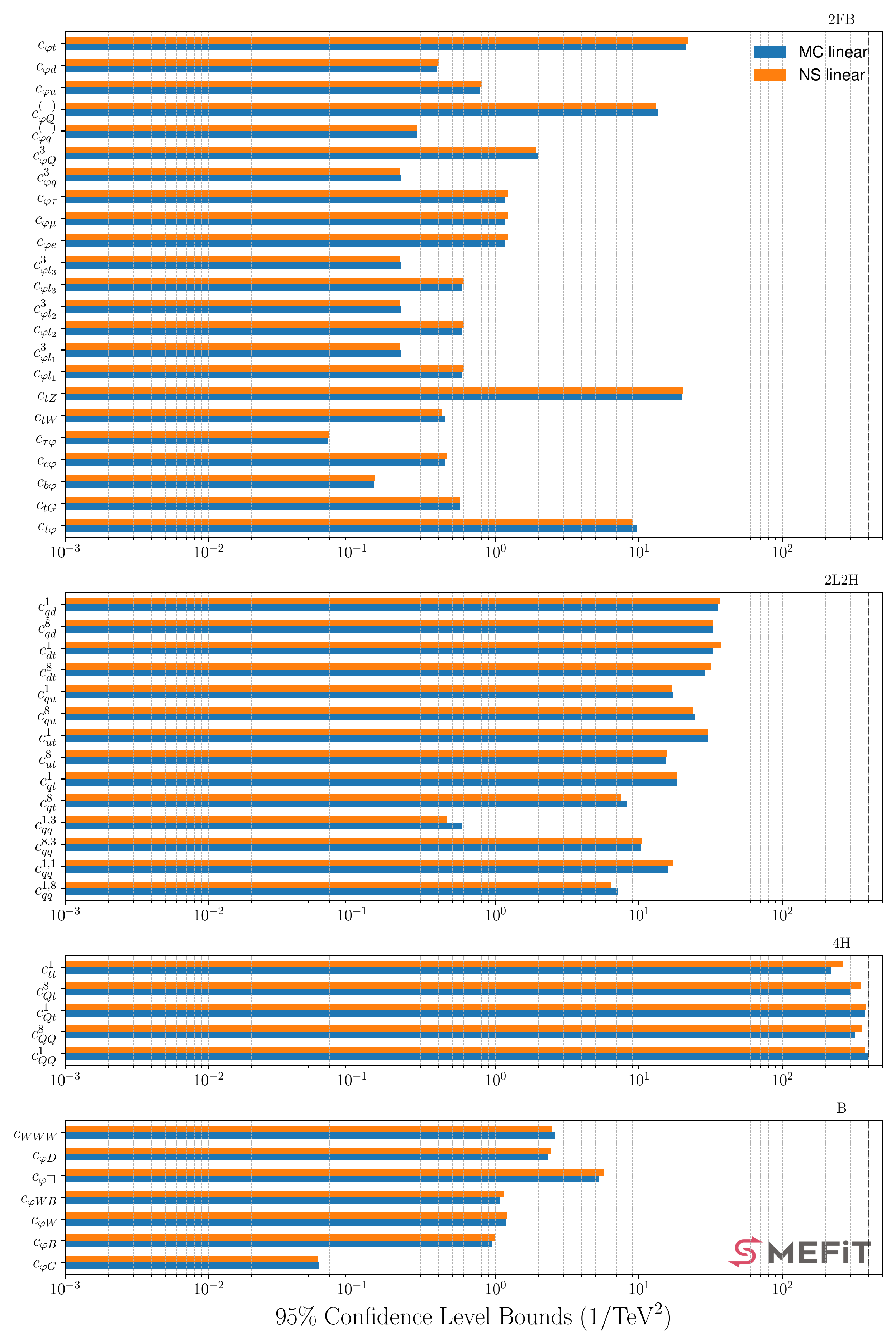}
    \includegraphics[width=0.49\textwidth]{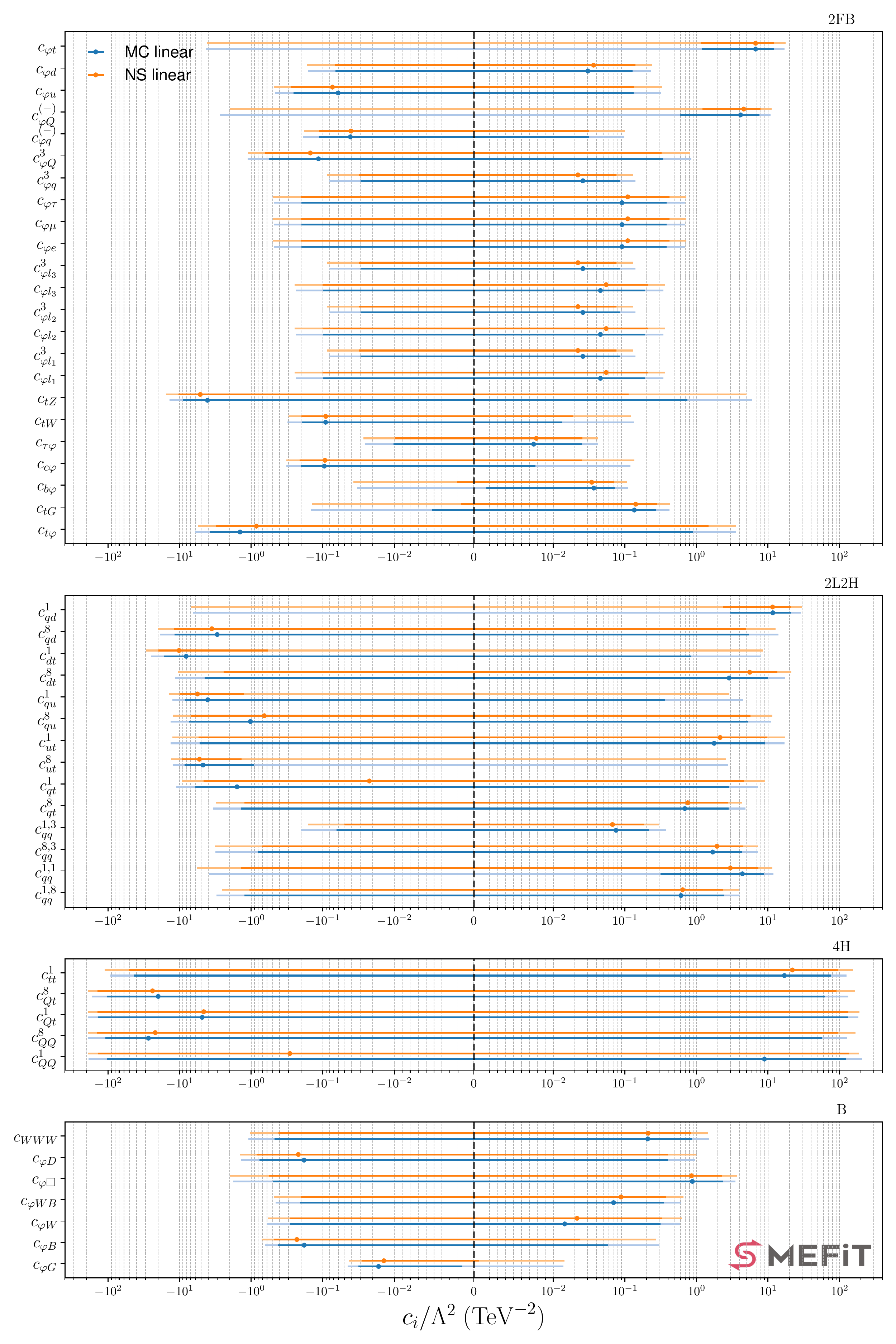}
    \caption{\small Results of global SMEFT fits with liner EFT corrections with the same
      inputs as in Fig.~\ref{fig:new_vs_old} obtained with either the NS or the MCfit methods.
      The left panel compares the magnitude of the 95\% CL intervals
      for $c_i/\Lambda^2$
      for the $n_{\rm op}=49$ Wilson coefficients
      considered in the analysis, while
      the right panel displays the median and 68\% and 95\% CL (thick and thin lines
      respectively) intervals in
      each case.
      Results are grouped by operator family: from top to bottom we show
      the two-fermion, two-light-two-heavy four-fermion,
      the four-heavy-fermion, and the purely bosonic operators.
    }
    \label{fig:report_MC_NS_linear}
\end{figure}

Fig.~\ref{fig:report_MC_NS_linear} presents the
comparison between linear EFT fits performed with NS and MCfit 
for the same data and theory settings as in global SMEFT fit of Fig.~\ref{fig:new_vs_old}.
The left panel compares the magnitude of the 95\% CL intervals
for $c_i/\Lambda^2$
for the $n_{\rm op}=49$ Wilson coefficients
considered in the analysis, while
the right panel displays the median and 68\% CL  and 95\% CL (thick and thin
respectively) intervals in each case.
Results are grouped by operator family: from top to bottom we show
the two-fermion, two-light-two-heavy four-fermion,
the four-heavy-fermion, and the purely bosonic operators.

In the case of the linear EFT fits, results obtained with NC and MCfit 
are essentially identical at the level of the (Gaussian) posterior distributions, 
though a large number of replicas $N_{\rm rep}$ is required in MCfit 
to achieve sufficiently smooth shapes of the distributions, 
with residual differences moderate and confined to poorly constrained operators,
such as the four-heavy top quark operators, that have a small contribution to the total
$\chi^2$.
Given that the NS and MCfit methods are orthogonal to each other, their
agreement constitutes a non-trivial cross-check of the robustness of the global SMEFT
analysis framework.
The  availability of such functionality is specially useful
to exclude that eventual deviations with respect to the  SM baseline
can be traced back to methodological limitations of the fitting framework.

On the other hand, when considering quadratic EFT fits,
the agreement between  NS and MCfit worsens for specific
operators.
This problem has been investigated in App.~E of~\cite{Kassabov:2023hbm},
where  analytical calculations are performed
for the posterior distributions in the case of single-parameter quadratic SMEFT fits,
finding that MCfit results may not reproduce the correct Bayesian posteriors
obtained from NS due to  spurious solutions related to cancellations between the linear and quadratic EFT terms.
This effect is most marked for observables where the quadratic EFT corrections dominate 
over the linear ones, and also whenever the SM cross-section overshoots sizably the central value 
of the experimental data.
Hence~\cite{Kassabov:2023hbm} finds that NS and MCfit will only coincide
at the quadratic level when all processes included in the fit are such that
quadratic EFT corrections are subdominant as compared to the linear ones.
However, for many of the observables considered in the global SMEFT fit, 
in particular those sensitive to the high energy tails accessible at the LHC, 
quadratic EFT effects are large and hence MCfit results may differ from
the NS posteriors.

%% file: sec-atlaseft.tex
\section{The ATLAS Higgs EFT analysis as case study}
\label{sec:atlaseft}

To illustrate the potential applications
of the {\sc\small SMEFiT} framework, we independently reproduce
the results of the ATLAS EFT interpretation presented in~\cite{ATL-PHYS-PUB-2022-037},
which  updates and supersedes previous ATLAS EFT
studies~\cite{ATLAS-CONF-2021-053, ATL-PHYS-PUB-2021-022}.
This analysis is
based on the combination of ATLAS measurements of Higgs production, diboson production,
and $Z$ production in vector boson fusion 
with legacy electroweak precision observables from LEP and SLC.

Our {\sc\small SMEFiT}-based reinterpretation is based on the same experimental data inputs, SM predictions,
EFT cross-section parametrisations, and operator basis rotations
as those used in~\cite{ATL-PHYS-PUB-2022-037}.
This information is  publicly available
for the linear $\mathcal{O}\lp \Lambda^{-2}\rp$ case: specifically,
the central values, uncertainties and correlations of the experimental measurements have been 
extracted from Tables~4, 9, 10 and Fig.~18 of~\cite{ATL-PHYS-PUB-2022-037};
the definition of the rotated fit basis in terms of the Warsaw basis from Fig. 13; and for the linear EFT
cross-sections
a prescription based on the public numbers provided in~\cite{ATL-PHYS-PUB-2022-037,ATLAS:2021vrm,ATLAS:2021ohb}
has been produced and used as input for the current analysis.
These public inputs are also made available in the {\sc\small SMEFiT} repository, in order
to facilitate the reproducibility of this benchmarking exercise.

The ATLAS EFT interpretation of~\cite{ATL-PHYS-PUB-2022-037}
is based on Higgs boson production cross-sections and decay measurements
carried out
 within the Simplified Template
Cross-Section (STXS) framework~\cite{Berger:2019wnu} from the Run II
dataset.
It also contains selected electroweak Run II measurements,
in particular diboson production in the $WW$, $WZ$, and $4\ell$ final states
as well as $Z$ production in vector-boson-fusion, $pp\to Z(\to \ell^+\ell^-)jj$.
Note that the  diboson $4\ell$ final state targets both on-shell $ZZ$ 
as well as off-shell Higgs boson production.
Table~\ref{tab:ATLAS_EW_data} summarizes the 
information associated to these ATLAS experimental inputs.

\begin{table}[h]
    \centering
    \footnotesize
    \input{LHCdata.tex}
    \vspace{0.3cm}
    \caption{The ATLAS measurements included in the EFT intepretation
      of~\cite{ATL-PHYS-PUB-2022-037}.
      For each dataset we provide specific details of the measurement, the
      integrated luminosity $\mathcal{L}$, and the corresponding publication reference.
      For Higgs measurements we display which production modes are being targeted
      in the analysis.
      For electroweak measurements we indicate the differential distribution
      included in the fit and the main acceptance cuts.
      In this table $\ell =e,\mu$ denotes a first- or second-generation charged lepton. 
    }
    \label{tab:STXS_data}  \label{tab:ATLAS_EW_data}
\end{table}

The ATLAS data listed in Table~\ref{tab:ATLAS_EW_data} are complemented by
the legacy LEP and SLC electroweak precision  observables (EWPO)
at the $Z$-pole from~\cite{ALEPH:2005ab},
 required to constrain directions in the SMEFT
parameter space not covered by LHC processes.
Specifically, the analysis of~\cite{ATL-PHYS-PUB-2022-037} considers
the inclusive cross-section
into hadrons $\sigma^0_{\text{had}}$, 
the ratio of partial decay widths $R_\ell^0$, $R_q^0$,
and the forward-backward asymmetries $A^{0,\ell}_{\rm fb}$,
$A^{0,q}_{\rm fb}$ where $q$ is measured separately for charm
and bottom quarks and $\ell$ is the average over leptons.
These EWPOs are defined as
\begin{align}
    \sigma^0_{\text{had}} = \frac{12\pi}{m^2_Z}\frac{\Gamma_{ee}\Gamma_{\text{had}}}{\Gamma^2_Z}\,,\quad
    R_\ell^0 = \frac{\Gamma_{\text{had}}}{\Gamma_{\ell\ell}}\,,\quad R_q^0 = \frac{\Gamma_{qq}}{\Gamma_{\text{had}}}\,,\quad
    A^0_{\rm fb} = \frac{N_F-N_B}{N_F+N_B}\,,
\end{align}
where $\Gamma_{Z}$, $\Gamma_{ee}$, $\Gamma_{\text{had}}$, $\Gamma_{\ell\ell}$ and $\Gamma_{qq}$ are the total and 
partial decay widths for the $Z$ boson and $q = c, b$, and
$N_F$ ($N_B$) indicates the number of events in which the final-state
fermion is produced in the forward (backward) direction.
Further details about the EWPO implementation can be found in~\cite{ATL-PHYS-PUB-2022-037}
and references therein.

Since the goal of this benchmarking exercise
is to carry out an independent validation of the results of~\cite{ATL-PHYS-PUB-2022-037}
using the same theory and data inputs but now with the \smefit code,
as indicated above
we take the SM and linear EFT cross-sections from the ATLAS note and parse them
into the \smefit format  adopting the same flavour assumptions for the fitting basis,
namely ${\rm U}(2)_q \times {\rm U}(2)_u \times {\rm U}(2)_d \times {\rm U}(3)_\ell \times {\rm U}(3)_e$.
CP conservation is assumed and Wilson coefficients are real-valued.
SM Higgs cross-sections 
are taken from the LHC Higgs WG~\cite{LHCHiggsCrossSectionWorkingGroup:2016ypw},
while LHC electroweak processes 
are computed using {\sc \small Sherpa2.2.2}~\cite{Sherpa:2019gpd},
{\sc \small Herwig7.1.5}~\cite{Bahr:2008pv}, and {\sc \small VBFNLO3.0.0}~\cite{Baglio:2011juf}
at NLO, matched to  {\sc \small Sherpa} and {\sc \small Pythia8}~\cite{Sjostrand:2014zea}
parton showers respectively.
Linear EFT cross-sections are computed with {\sc \small MadGraph5\_aMC@NLO }~\cite{Alwall:2011uj}
and {\sc \small SMEFTsim}~\cite{Brivio:2017btx, Brivio:2020onw},
except for  loop-induced processes in the SM such as 
 $gg\rightarrow h$ and $gg \rightarrow Zh$ where
{\sc \small SMEFT@NLO}~\cite{Degrande:2020evl} is used for the calculation of 1-loop
QCD effects.
An analytical computation 
with NLO accuracy in QED~\cite{Dawson:2018liq} is used for $H\rightarrow \gamma\gamma$.
SMEFT propagator effects impacting the mass and width
of intermediate particles are computed using {\sc \small SMEFTsim}.
The {\sc \small MadGraph5\_aMC@NLO+Pythia8} predictions
are supplemented with bin-by-bin $K$-factors to account for higher-order QCD and electroweak corrections. 
Theory predictions for  EWPOs in the SM and the SMEFT follow~\cite{Corbett:2021eux}
adapted to the flavour assumptions of~\cite{ATL-PHYS-PUB-2022-037}.

The SMEFT predictions in the Warsaw basis for the processes
entering the analysis of~\cite{ATL-PHYS-PUB-2022-037} depend on $n_{\rm op}=62$ independent Wilson coefficients.
However, at the linear level only a subset of directions can be constrained from the input measurements,
with the other linearly independent combinations leading to flat directions in the likelihood function $\mathcal{L}({\boldsymbol c})$. 
Numerical minimizers such as that used in~\cite{ATL-PHYS-PUB-2022-037} can only work with problems where there exists
    a point solution in the parameter space, unless all strictly flat directions are removed.
    For this reason,
the ATLAS fit of~\cite{ATL-PHYS-PUB-2022-037} is carried
out not in the Warsaw basis but in a rotated basis, corresponding to the directions
with the highest variability as determined by a principal component analysis 
of the matrix
\begin{align}
  \mathcal{H}_{i,j} \equiv \sum_{m,n=1}^{n_{\rm dat}}\sigma_{{\rm eft},i,m} ({\rm cov}^{-1})_{m,n}  \sigma_{{\rm eft},j,n}\,,
\end{align}
which can be identified with the Fisher information matrix,
Eq.~(\ref{eq:FisherDef}).
The PCA defines the rotation matrix $R^{\rm{(W\to A)}}_{ij}$ that implements this basis transformation
\begin{align}
  \label{eq:PCA-rotatedbasis}
  c^{\text{(A)}}_{i} = \sum_{j=1}^{n_{\rm op}=62} R^{\rm{(W\to A)}}_{ij} c^{\text{(W)}}_{j}\,, \qquad R^{\rm{(A\to W)}}= \lp R^{\rm{(W\to A)}}\rp^{-1} \, ,
\end{align}
where $(A)$ indicates the ATLAS fit basis and $(W)$ the Warsaw basis, see Sect.~5.2 and Fig.~8 
of~\cite{ATL-PHYS-PUB-2022-037} for the explicit definitions.
The ATLAS analysis is then performed in terms of the 28 PCA eigenvectors ${\boldsymbol c}^{(A)}$ with the highest variability,
with the remaining 34 linear combinations set to zero.
Below we demonstrate how the results of~\cite{ATL-PHYS-PUB-2022-037} are also reproduced
when the fit is carried out directly in the original 62-dimensional
Warsaw basis ${\boldsymbol c}^{(W)}$  rather than in the PCA-rotated basis.

\begin{figure}[t]
    \center
    \includegraphics[scale=0.35]{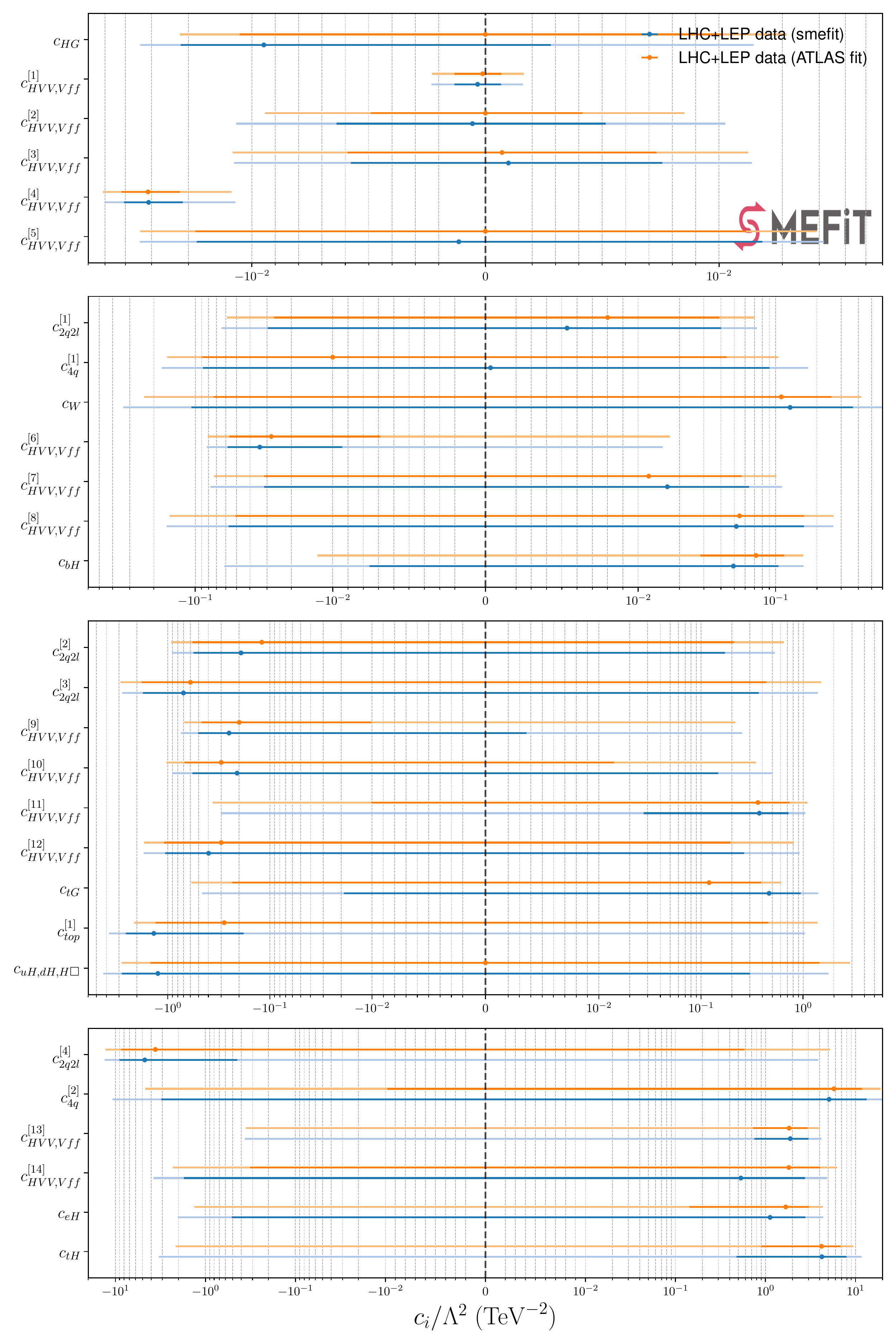}
    \includegraphics[scale=0.35]{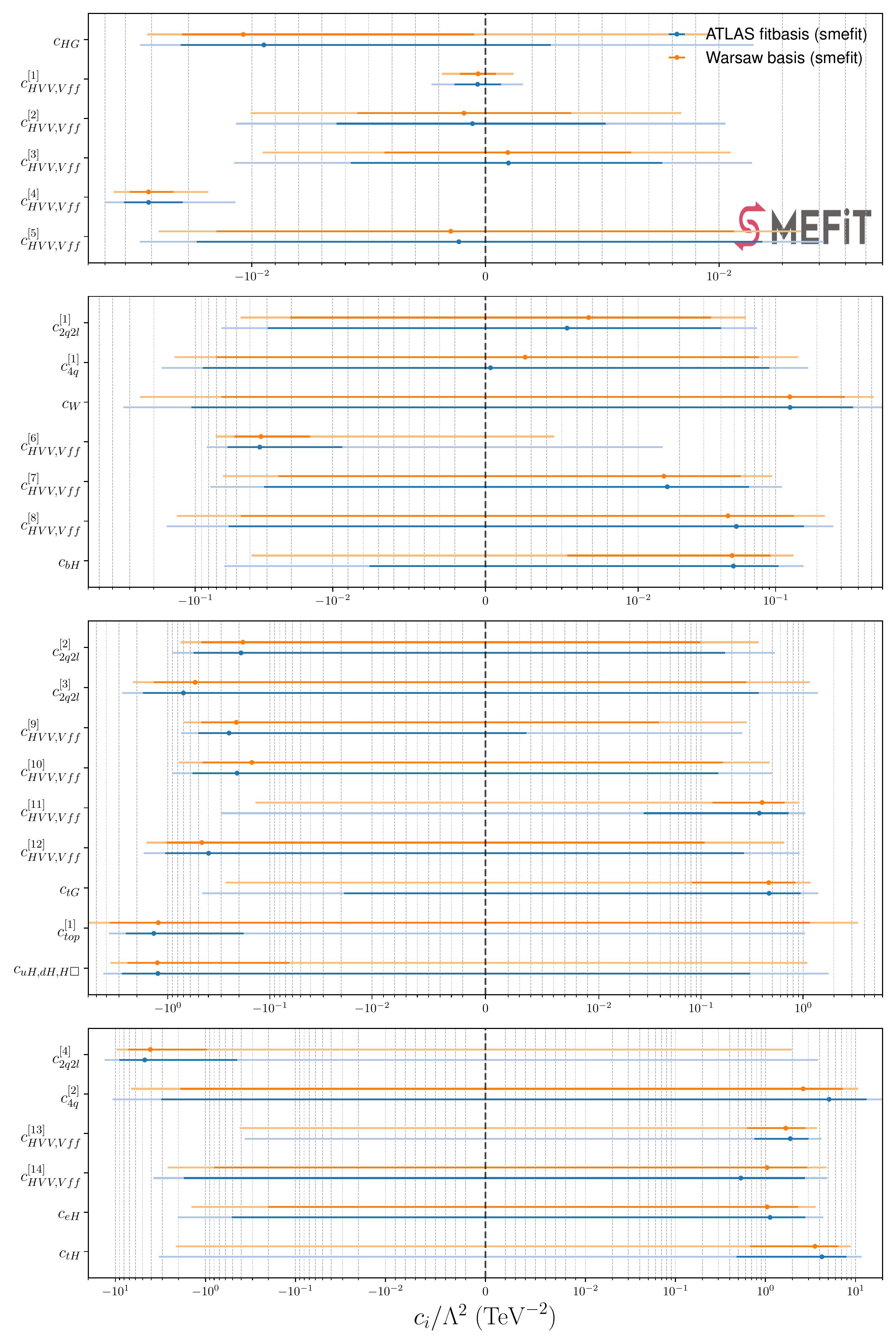}
    \caption{\small Left panel: comparison of the ATLAS EFT fit results from~\cite{ATL-PHYS-PUB-2022-037} with the corresponding
      results based on the \smefit code when the same theory, data inputs,
      and fitting basis are adopted.
      The dark and pale lines represent the $68\%$ and $95\%$~CL intervals respectively.
      Right panel: same as the left panel when the results of the \smefit 
      analysis of ATLAS and LEP data obtained in the ATLAS fit basis  are compared to those obtained
      when fitting directly in the relevant
      $n_{\rm op}=62$ dimensional subset of the Warsaw basis and then
      rotated
      to the ATLAS fit basis.
    }
    \label{fig:atlas_comparisons}
\end{figure}

The left panel of Fig.~\ref{fig:atlas_comparisons} compares the ATLAS EFT fit results
from~\cite{ATL-PHYS-PUB-2022-037} with the corresponding
results obtained with the \smefit code when the same theory, data inputs,
and fitting basis are adopted.
The outcome of the ATLAS analysis is provided in~\cite{ATL-PHYS-PUB-2022-037}
 both for the full likelihood and for a
simplified multi-Gaussian likelihood;  here we consider the latter to ensure a consistent comparison
with the \smefit results.
The dark and pale lines represent the $68\%$ and $95\%$~CL intervals respectively.
Since the EFT calculations include only linear cross-sections, the resulting posteriors
are by construction Gaussian.
In both cases, the fits have been carried out in the 28-dimensional
PCA-rotated basis ${\boldsymbol c}^{(A)}$  defined by Eq.~(\ref{eq:PCA-rotatedbasis}).
The \smefit output corresponds to Nested Sampling, though equivalent results are obtained with MCfit.

\begin{figure}[t]
    \center
    \includegraphics[scale=0.5]{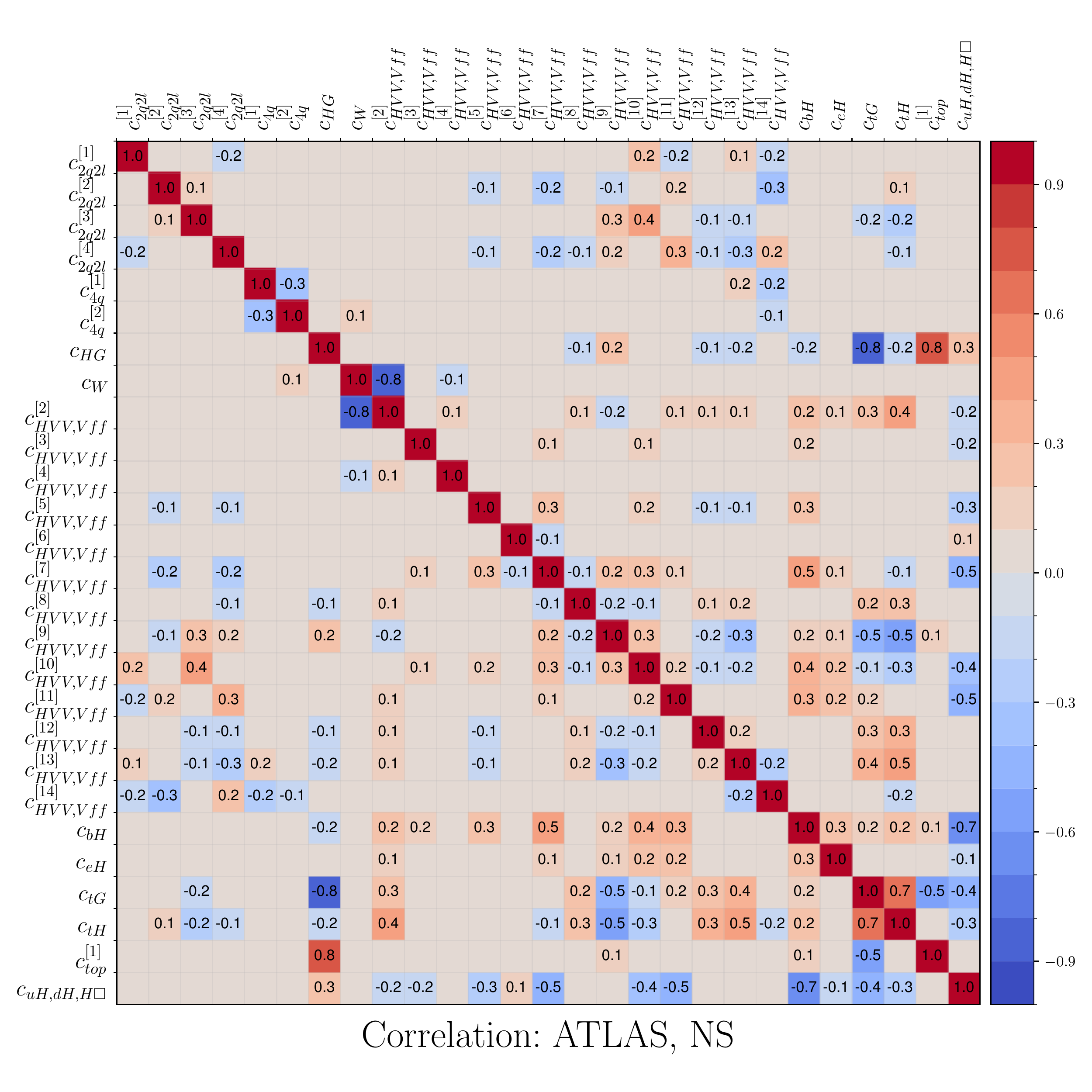}
    \caption{The correlation coefficients obtained in the \smefit analysis
      reproducing~\cite{ATL-PHYS-PUB-2022-037} in the ATLAS fit basis.
      We do not display the numerical values of the correlation matrix
      entries with $|\rho_{ij}|<0.1$.
      }
    \label{fig:correlations_LHC+LEP}
\end{figure}

Inspection of Fig.~\ref{fig:atlas_comparisons} confirms that
good agreement is obtained both in terms of central values and of the uncertainties of
the fitted Wilson coefficients.
Furthermore, similar agreement is obtained for the correlations $\rho_{ij}$ between EFT coefficients,
displayed in Fig.~\ref{fig:correlations_LHC+LEP} in the PCA-rotated  basis,
as can be verified by comparing with the results from~\cite{ATL-PHYS-PUB-2022-037}.
The fact that the entries of correlation matrix displayed in Fig.~\ref{fig:correlations_LHC+LEP}
are typically small, with  few exceptions, is a consequence of using a rotated fit
basis which by construction
reduces the correlations between fitted degrees of freedom.

As discussed in Sect.~\ref{sec:framework}, within the \smefit framework
it is possible to rotate from the Warsaw basis to any user-defined
operator basis.
In addition, the user can choose to automatically rotate to a fitting basis
defined by the  principal components
of the Fisher information via Eq.~(\ref{eq:PCA-rotatedbasis-sec2}),
indicating the threshold restricting the kept singular values.
While being numerically less efficient,
the presence of flat directions does not represent a bottleneck in
\smefit when the Nested Sampling strategy is adopted.
One can therefore
combine these two functionalities to repeat the EFT interpretation displayed in the left
panel of Fig.~\ref{fig:atlas_comparisons} now in the original 62-dimensional Warsaw basis ${\boldsymbol c}^{(W)}$,
rather than in the 28-dimensional PCA-rotated basis ${\boldsymbol c}^{(A)}$.
Afterwards, one can
use Eq.~(\ref{eq:PCA-rotatedbasis}) to rotate the obtained posterior distributions
from the Warsaw to the ATLAS fit basis and assess whether or not fit results
are indeed independent of the basis choice.

The outcome of this exercise is reported in the right panel of Fig.~\ref{fig:atlas_comparisons},
and compared to the  \smefit results obtained
when using the PCA-rotated fitting basis ${\boldsymbol c}^{(A)}$.
Excellent agreement is also found in this case,
 demonstrating the basis independence of the SMEFT
interpretation of the dataset entering~\cite{ATL-PHYS-PUB-2022-037}.
Similar basis stability tests could
be carried out with any other basis related to the Warsaw by a unitary transformation.
We note that this property holds true only when the rotation
Eq.~(\ref{eq:PCA-rotatedbasis}) is applied
sample by sample (replica by replica) in NS (MCfit), rather than at the level
of mean values and CL intervals.
Posterior distributions are also basis independent, and as expected the 34 principal
components
excluded from ${\boldsymbol c}^{(A)}$ display posteriors which are flat or quasi-flat.
This comparison hence confirms
the robustness of our EFT analysis method in the presence of flat 
directions in the parameter space.

The benchmarking exercise displayed in Fig.~\ref{fig:atlas_comparisons}
illustrates how while specific choices of operator bases may
be preferred in terms of numerical efficiency or clarity of the physical
interpretation, ultimately the EFT fit results should be independent of this choice.
This feature is specially relevant to compare results obtained
by different groups, which usually adopt different fitting bases.

%% file: LHCdata.tex
\small
\renewcommand{\arraystretch}{1.70}
\begin{tabular}{llcc}
 \toprule
 Process & Details  &   $\qquad \mathcal{L}\left[\text{fb}^{-1}\right]\qquad $ & Ref.  \\   
 \hline
 $pp\to h \rightarrow \gamma\gamma$ & ggF, VBF, $Wh$, $Zh$, $t\bar{t}h$, $th$ & 139 & \cite{ATLAS-CONF-2020-026} \\
 $pp\to h \rightarrow ZZ^*$ &  ggF, VBF, $Wh$, $Zh$, $t\bar{t}h$(4$\ell$) & 139 & \cite{ATLAS:2020rej} \\
 $pp\to h \rightarrow WW^*$ &  ggF, VBF                          & 139 & \cite{ATLAS-CONF-2021-014} \\
 $pp\to h \rightarrow \tau\tau$ &  ggF, VBF, $Wh$, $Zh$, $t\bar{t}h$($\tau_{\text{had}}\tau_{\text{had}}$) & 139 & \cite{ATLAS-CONF-2021-044} \\ 
 \hline 
  & $Wh$, $Zh$ & 139 & \cite{ATLAS:2020fcp, ATLAS:2020jwz, ATLAS-CONF-2021-051} \\ 
$pp \to h \rightarrow b\bar{b}$ & VBF & 126 & \cite{ATLAS:2020bhl} \\ 
& $t\bar{t}h$ & 139 & \cite{ATLAS:2021qou} \\
\hline
 $pp \rightarrow e^{\pm}\nu\mu^{\mp}\nu$ & $p_T^{\text{lead. lep.}}$ $\left( m_{\ell\ell} > 55\, \text{GeV}\,, p_T^{\text{jet}}< 35\, \text{GeV}\rp $   & 36 & \cite{ATLAS:2019rob} \\
 $pp \rightarrow \ell^{\pm}\nu \ell^+ \ell^-$ & $m_T^{WZ}$  $\lp m_{\ell\ell} \in \left(81, 101\right)\,\text{GeV}\rp $  & 36 & \cite{ATLAS:2019bsc} \\
 $pp \rightarrow \ell^+\ell^-\ell^+\ell^-$ & $m_{ZZ}$ $\lp m_{4\ell} > 180\,\text{GeV}\rp $   & 139 & \cite{ATLAS:2021kog} \\
 $pp \rightarrow \ell^+\ell^-jj$ &  $\Delta\Phi_{jj}$ $\lp m_{jj}> 1000\, \text{GeV}\,,\,m_{\ell\ell} \in \left(81,101\right)\,\text{GeV}\rp $  & 139 & \cite{ATLAS:2020nzk} \\
 \hline
\end{tabular}

%% file: sec-summary.tex
\section{Summary and outlook}
\label{sec:summary}

In this work we have presented the open source
\smefit package, summarised its main functionalities,
demonstrated how it can be used to reproduce the outcome
of the global analysis of~\cite{Ethier:2021bye},
and independently reproduced the ATLAS EFT interpretation of LHC and LEP data
from~\cite{ATL-PHYS-PUB-2022-037} to highlight some of its possible applications.
We have deliberately kept to a minimum the technical details, for instance
concerning the format of the theory and data tables, and pointed the reader
to the developing online documentation for further information.
The features outlined in this paper represent only a snapshot
of the full code capabilities, and in particular do not cover the extensive post-analysis
visualization tools and statistical diagnosis methods provided with the code.
While we have not discussed code performance, this is currently not a limiting factor
for the analyses considered, with the global EFT fit (ATLAS EFT fit in Warsaw basis) 
in Fig.~\ref{fig:new_vs_old} (Fig.~\ref{fig:atlas_comparisons}) taking around 
2 (8) hours running on 24 cores.

This updated \smefit framework will be the stepping stone
making possible the realisation of a number of ongoing projects related to
global interpretations of particle physics data in the SMEFT.
To begin with, the implementation and (partial) automation of the matching between the
SMEFT and UV-complete scenarios at the fit level,
in a way that upon a choice of UV model, \smefit returns the posterior distributions
in the space of UV theory parameters such as heavy particle masses and couplings.
This functionality is enabled by the \smefit flexibility in
imposing arbitrary restrictions between the Wilson coefficients,
and will lead to the option of carrying out directly the fits in terms of the UV parameters,
rather than in terms of the EFT coefficients.
Second, to assess the impact in the global SMEFT fit of improved theory calculations
such as the inclusion of renormalisation group running effects~\cite{Aoude:2022aro}
and of electroweak corrections in high-energy observables.
Third, to carry out projections quantifying the reach in the SMEFT parameter space~\cite{deBlas:2022ofj,deBlas:2019rxi} of future lepton-lepton,
lepton-hadron, and hadron-hadron colliders when these measurements are added on top of a state-of-the-art
global fit.
Fourth, extending the EFT determination to novel types of measurements beyond those based
on a multi-Gaussian statistical model, such
as the unbinned multivariate observables presented in~\cite{GomezAmbrosio:2022mpm}.
Fifth, to validate related efforts such as the {\sc\small SimuNET} technique~\cite{Iranipour:2022iak}
developed to perform a simultaneous determination of the PDFs and EFT coefficients~\cite{Carrazza:2019sec,
  Greljo:2021kvv},
which should reduce to the \smefit outcome in the fixed-PDF case for the same choice of
theory and experimental inputs.

In addition to  physics-motivated developments such as those outlined above, we also plan to further
improve the statistical framework underlying \smefit and expand the visualization
and analysis post-processing tools provided.
One possible direction would be to implement new avenues to carry out parameter inference,
such as the ML-assisted simulation-based inference method proposed in~\cite{Miller:2020hua}, as well
as a broader range of optimisers for MCfit such as those studied in the benchmark
comparison of~\cite{DarkMachinesHighDimensionalSamplingGroup:2021wkt}.
It would also be advantageous to apply complementary methods to determine the more
and less constrained directions in the parameter space.
In particular, one could extend
the linear PCA analysis to non-linear algorithms relevant to the case where the  quadratic
EFT corrections become sizable, such as with t-Distributed Stochastic Neighbor Embedding (t-SNE).
Finally, one would like to run \smefit in hardware accelerators such
as  multi graphics processing units (GPUs), leading to a further speed up
of the code similar to that reported for PDF interpolations and event generators~\cite{Carrazza:2021gpx,Carrazza:2020qwu}.

The availability of this framework provides the SMEFT community with a new toolbox for all kinds
of EFT interpretations of experimental data, with its modular structure facilitating
the extension to other datasets and process types, updated theory calculations,
and eventually its application to other EFTs such as the Higgs EFT.
\smefit will also streamline the comparisons and benchmarking between EFT determinations
carried out by different groups, as the ATLAS analysis illustrates, and could be adopted
by the  experimental collaborations in order to cross-check
the results obtained in their own frameworks.

\paragraph{Acknowledgments.}
We are grateful to Jaco ter Hoeve for testing the current
version and for implementing the two-parameter contour feature in the fit report,
and to Samuel van Beek, Jacob J. Ethier, Nathan.~P.~Hartland, Emanuele R. Nocera, and Emma Slade for 
their contributions to previous versions of the {\sc\small SMEFiT} framework.
We thank Fabio Maltoni, Luca Mantani, Alejo Rossia, and Eleni Vryonidou for
their inputs and suggestions
concerning the {\sc\small SMEFiT} functionalities and
the choice of formats for the data and theory tables.
We are grateful to Rahul Balasubramanian, Lydia Brenner,  Oliver Rieger, Wouter Verkerke, and
Andrea Visibile for useful discussions about the ATLAS EFT Higgs analyses.

%% file: smefit-code.bbl
\providecommand{\href}[2]{#2}\begingroup\raggedright\begin{thebibliography}{10}

\bibitem{Weinberg:1979sa}
S.~Weinberg, {\it {Baryon and Lepton Nonconserving Processes}},  {\em Phys.
  Rev. Lett.} {\bf 43} (1979) 1566--1570.

\bibitem{Buchmuller:1985jz}
W.~Buchmuller and D.~Wyler, {\it {Effective Lagrangian Analysis of New
  Interactions and Flavor Conservation}},  {\em Nucl. Phys.} {\bf B268} (1986)
  621--653.

\bibitem{Grzadkowski:2010es}
B.~Grzadkowski, M.~Iskrzynski, M.~Misiak, and J.~Rosiek, {\it {Dimension-Six
  Terms in the Standard Model Lagrangian}},  {\em JHEP} {\bf 10} (2010) 085,
  [\href{http://arxiv.org/abs/1008.4884}{{\tt arXiv:1008.4884}}].

\bibitem{Hartland:2019bjb}
N.~P. Hartland, F.~Maltoni, E.~R. Nocera, J.~Rojo, E.~Slade, E.~Vryonidou, and
  C.~Zhang, {\it {A Monte Carlo global analysis of the Standard Model Effective
  Field Theory: the top quark sector}},  {\em JHEP} {\bf 04} (2019) 100,
  [\href{http://arxiv.org/abs/1901.05965}{{\tt arXiv:1901.05965}}].

\bibitem{Brivio:2019ius}
I.~Brivio, S.~Bruggisser, F.~Maltoni, R.~Moutafis, T.~Plehn, E.~Vryonidou,
  S.~Westhoff, and C.~Zhang, {\it {O new physics, where art thou? A global
  search in the top sector}},  {\em JHEP} {\bf 02} (2020) 131,
  [\href{http://arxiv.org/abs/1910.03606}{{\tt arXiv:1910.03606}}].

\bibitem{Biekotter:2018rhp}
A.~Biekötter, T.~Corbett, and T.~Plehn, {\it {The Gauge-Higgs Legacy of the
  LHC Run II}},  {\em SciPost Phys.} {\bf 6} (2019) 064,
  [\href{http://arxiv.org/abs/1812.07587}{{\tt arXiv:1812.07587}}].

\bibitem{Ellis:2018gqa}
J.~Ellis, C.~W. Murphy, V.~Sanz, and T.~You, {\it {Updated Global SMEFT Fit to
  Higgs, Diboson and Electroweak Data}},  {\em JHEP} {\bf 06} (2018) 146,
  [\href{http://arxiv.org/abs/1803.03252}{{\tt arXiv:1803.03252}}].

\bibitem{Almeida:2018cld}
E.~da~Silva~Almeida, A.~Alves, N.~Rosa~Agostinho, O.~J. \'Eboli, and
  M.~Gonzalez-Garcia, {\it {Electroweak Sector Under Scrutiny: A Combined
  Analysis of LHC and Electroweak Precision Data}},  {\em Phys. Rev. D} {\bf
  99} (2019), no.~3 033001, [\href{http://arxiv.org/abs/1812.01009}{{\tt
  arXiv:1812.01009}}].

\bibitem{Aebischer:2018iyb}
J.~Aebischer, J.~Kumar, P.~Stangl, and D.~M. Straub, {\it {A Global Likelihood
  for Precision Constraints and Flavour Anomalies}},  {\em Eur. Phys. J. C}
  {\bf 79} (2019), no.~6 509, [\href{http://arxiv.org/abs/1810.07698}{{\tt
  arXiv:1810.07698}}].

\bibitem{Ellis:2020unq}
J.~Ellis, M.~Madigan, K.~Mimasu, V.~Sanz, and T.~You, {\it {Top, Higgs, Diboson
  and Electroweak Fit to the Standard Model Effective Field Theory}},  {\em
  JHEP} {\bf 04} (2021) 279, [\href{http://arxiv.org/abs/2012.02779}{{\tt
  arXiv:2012.02779}}].

\bibitem{Bissmann:2020mfi}
S.~Bi\ss{}mann, C.~Grunwald, G.~Hiller, and K.~Kr\"oninger, {\it {Top and
  Beauty synergies in SMEFT-fits at present and future colliders}},  {\em JHEP}
  {\bf 06} (2021) 010, [\href{http://arxiv.org/abs/2012.10456}{{\tt
  arXiv:2012.10456}}].

\bibitem{Bruggisser:2021duo}
S.~Bruggisser, R.~Sch\"afer, D.~van Dyk, and S.~Westhoff, {\it {The Flavor of
  UV Physics}},  {\em JHEP} {\bf 05} (2021) 257,
  [\href{http://arxiv.org/abs/2101.07273}{{\tt arXiv:2101.07273}}].

\bibitem{Bruggisser:2022rhb}
S.~Bruggisser, D.~van Dyk, and S.~Westhoff, {\it {Resolving the Flavor
  Structure in the MFV-SMEFT}},  \href{http://arxiv.org/abs/2212.02532}{{\tt
  arXiv:2212.02532}}.

\bibitem{Ethier:2021bye}
{\bf SMEFiT} Collaboration, J.~J. Ethier, G.~Magni, F.~Maltoni, L.~Mantani,
  E.~R. Nocera, J.~Rojo, E.~Slade, E.~Vryonidou, and C.~Zhang, {\it {Combined
  SMEFT interpretation of Higgs, diboson, and top quark data from the LHC}},
  {\em JHEP} {\bf 11} (2021) 089, [\href{http://arxiv.org/abs/2105.00006}{{\tt
  arXiv:2105.00006}}].

\bibitem{DeBlas:2019ehy}
J.~De~Blas et~al., {\it {$\texttt{HEPfit}$: a code for the combination of
  indirect and direct constraints on high energy physics models}},  {\em Eur.
  Phys. J. C} {\bf 80} (2020), no.~5 456,
  [\href{http://arxiv.org/abs/1910.14012}{{\tt arXiv:1910.14012}}].

\bibitem{Castro:2016jjv}
N.~Castro, J.~Erdmann, C.~Grunwald, K.~Kröninger, and N.-A. Rosien, {\it
  {EFTfitter---A tool for interpreting measurements in the context of effective
  field theories}},  {\em Eur. Phys. J.} {\bf C76} (2016), no.~8 432,
  [\href{http://arxiv.org/abs/1605.05585}{{\tt arXiv:1605.05585}}].

\bibitem{Ethier:2021ydt}
J.~J. Ethier, R.~Gomez-Ambrosio, G.~Magni, and J.~Rojo, {\it {SMEFT analysis of
  vector boson scattering and diboson data from the LHC Run II}},  {\em Eur.
  Phys. J. C} {\bf 81} (2021), no.~6 560,
  [\href{http://arxiv.org/abs/2101.03180}{{\tt arXiv:2101.03180}}].

\bibitem{Degrande:2020evl}
C.~Degrande, G.~Durieux, F.~Maltoni, K.~Mimasu, E.~Vryonidou, and C.~Zhang,
  {\it {Automated one-loop computations in the standard model effective field
  theory}},  {\em Phys. Rev. D} {\bf 103} (2021), no.~9 096024,
  [\href{http://arxiv.org/abs/2008.11743}{{\tt arXiv:2008.11743}}].

\bibitem{vanBeek:2019evb}
S.~van Beek, E.~R. Nocera, J.~Rojo, and E.~Slade, {\it {Constraining the SMEFT
  with Bayesian reweighting}},  {\em SciPost Phys.} {\bf 7} (2019), no.~5 070,
  [\href{http://arxiv.org/abs/1906.05296}{{\tt arXiv:1906.05296}}].

\bibitem{Castro:2022zpq}
N.~Castro, K.~Cranmer, A.~V. Gritsan, J.~Howarth, G.~Magni, K.~Mimasu,
  J.~Rojotwoaff, J.~Roskes, E.~Vryonidou, and T.~You, {\it {LHC EFT WG Report:
  Experimental Measurements and Observables}},
  \href{http://arxiv.org/abs/2211.08353}{{\tt arXiv:2211.08353}}.

\bibitem{ATL-PHYS-PUB-2022-037}
{\bf ATLAS} Collaboration, {\it {Combined effective field theory interpretation
  of Higgs boson and weak boson production and decay with ATLAS data and
  electroweak precision observables}},  tech. rep., CERN, Geneva, 2022.
\newblock All figures including auxiliary figures are available at
  https://atlas.web.cern.ch/Atlas/GROUPS/PHYSICS/PUBNOTES/ATL-PHYS-PUB-2022-037.

\bibitem{GomezAmbrosio:2022mpm}
R.~Gomez~Ambrosio, J.~ter Hoeve, M.~Madigan, J.~Rojo, and V.~Sanz, {\it
  {Unbinned multivariate observables for global SMEFT analyses from machine
  learning}},  \href{http://arxiv.org/abs/2211.02058}{{\tt arXiv:2211.02058}}.

\bibitem{Feroz:2013hea}
F.~Feroz, M.~P. Hobson, E.~Cameron, and A.~N. Pettitt, {\it {Importance Nested
  Sampling and the MultiNest Algorithm}},
  \href{http://arxiv.org/abs/1306.2144}{{\tt arXiv:1306.2144}}.

\bibitem{Feroz:2007kg}
F.~Feroz and M.~Hobson, {\it {Multimodal nested sampling: an efficient and
  robust alternative to MCMC methods for astronomical data analysis}},  {\em
  Mon. Not. Roy. Astron. Soc.} {\bf 384} (2008) 449,
  [\href{http://arxiv.org/abs/0704.3704}{{\tt arXiv:0704.3704}}].

\bibitem{DelDebbio:2004xtd}
{\bf NNPDF} Collaboration, L.~Del~Debbio, S.~Forte, J.~I. Latorre, A.~Piccione,
  and J.~Rojo, {\it {Unbiased determination of the proton structure function
  F(2)**p with faithful uncertainty estimation}},  {\em JHEP} {\bf 03} (2005)
  080, [\href{http://arxiv.org/abs/hep-ph/0501067}{{\tt hep-ph/0501067}}].

\bibitem{Ball:2008by}
{\bf The NNPDF} Collaboration, R.~D. Ball et~al., {\it {A determination of
  parton distributions with faithful uncertainty estimation}},  {\em Nucl.
  Phys.} {\bf B809} (2009) 1--63, [\href{http://arxiv.org/abs/0808.1231}{{\tt
  arXiv:0808.1231}}].

\bibitem{cmaes}
N.~Hansen and A.~Ostermeier, {\it Completely derandomized self-adaptation in
  evolution strategies},  {\em Evolutionary Computation} {\bf 9} (2001), no.~2
  159--195,
  [\href{http://arxiv.org/abs/https://doi.org/10.1162/106365601750190398}{{\tt
  https://doi.org/10.1162/106365601750190398}}].

\bibitem{Bertone:2017tyb}
{\bf NNPDF} Collaboration, V.~Bertone, S.~Carrazza, N.~P. Hartland, E.~R.
  Nocera, and J.~Rojo, {\it {A determination of the fragmentation functions of
  pions, kaons, and protons with faithful uncertainties}},  {\em Eur. Phys. J.}
  {\bf C77} (2017), no.~8 516, [\href{http://arxiv.org/abs/1706.07049}{{\tt
  arXiv:1706.07049}}].

\bibitem{Byrd1999AnIP}
R.~H. Byrd, M.~E. Hribar, and J.~Nocedal, {\it An interior point algorithm for
  large-scale nonlinear programming},  {\em SIAM J. Optim.} {\bf 9} (1999)
  877--900.

\bibitem{1997PhLA..233..216X}
Y.~{Xiang}, D.~Y. {Sun}, W.~{Fan}, and X.~G. {Gong}, {\it {Generalized
  simulated annealing algorithm and its application to the Thomson model}},
  {\em Physics Letters A} {\bf 233} (Feb., 1997) 216--220.

\bibitem{AbdulKhalek:2019bux}
{\bf NNPDF} Collaboration, R.~Abdul~Khalek et~al., {\it {A first determination
  of parton distributions with theoretical uncertainties}},  {\em Eur. Phys.
  J.} {\bf C} (2019) 79:838, [\href{http://arxiv.org/abs/1905.04311}{{\tt
  arXiv:1905.04311}}].

\bibitem{AbdulKhalek:2019ihb}
{\bf NNPDF} Collaboration, R.~Abdul~Khalek et~al., {\it {Parton Distributions
  with Theory Uncertainties: General Formalism and First Phenomenological
  Studies}},  {\em Eur. Phys. J. C} {\bf 79} (2019), no.~11 931,
  [\href{http://arxiv.org/abs/1906.10698}{{\tt arXiv:1906.10698}}].

\bibitem{AguilarSaavedra:2018nen}
D.~Barducci et~al., {\it {Interpreting top-quark LHC measurements in the
  standard-model effective field theory}},
  \href{http://arxiv.org/abs/1802.07237}{{\tt arXiv:1802.07237}}.

\bibitem{Ball:2011gg}
R.~D. Ball, V.~Bertone, F.~Cerutti, L.~Del~Debbio, S.~Forte, et~al., {\it
  {Reweighting and Unweighting of Parton Distributions and the LHC W lepton
  asymmetry data}},  {\em Nucl.Phys.} {\bf B855} (2012) 608--638,
  [\href{http://arxiv.org/abs/1108.1758}{{\tt arXiv:1108.1758}}].

\bibitem{Kassabov:2023hbm}
Z.~Kassabov, M.~Madigan, L.~Mantani, J.~Moore, M.~M. Alvarado, J.~Rojo, and
  M.~Ubiali, {\it {The top quark legacy of the LHC Run II for PDF and SMEFT
  analyses}},  \href{http://arxiv.org/abs/2303.06159}{{\tt arXiv:2303.06159}}.

\bibitem{ATLAS-CONF-2021-053}
{\bf ATLAS} Collaboration, {\it {Combined measurements of Higgs boson
  production and decay using up to $139$ fb$^{-1}$ of proton-proton collision
  data at $\sqrt{s}= 13$ TeV collected with the ATLAS experiment}},  tech.
  rep., CERN, Geneva, 2021.
\newblock All figures including auxiliary figures are available at
  https://atlas.web.cern.ch/Atlas/GROUPS/PHYSICS/CONFNOTES/ATLAS-CONF-2021-053.

\bibitem{ATL-PHYS-PUB-2021-022}
{\bf ATLAS} Collaboration, {\it {Combined effective field theory interpretation
  of differential cross-sections measurements of WW, WZ, 4l, and
  Z-plus-two-jets production using ATLAS data}},  tech. rep., CERN, Geneva,
  2021.
\newblock All figures including auxiliary figures are available at
  https://atlas.web.cern.ch/Atlas/GROUPS/PHYSICS/PUBNOTES/ATL-PHYS-PUB-2021-022.

\bibitem{ATLAS:2021vrm}
{\bf ATLAS} Collaboration, {\it {Combined measurements of Higgs boson
  production and decay using up to $139$ fb$^{-1}$ of proton-proton collision
  data at $\sqrt{s}= 13$ TeV collected with the ATLAS experiment}}, .

\bibitem{ATLAS:2021ohb}
{\bf ATLAS} Collaboration, {\it {Combined effective field theory interpretation
  of differential cross-sections measurements of $WW$, $WZ$, 4$\ell$, and
  $Z$-plus-two-jets production using ATLAS data}}, .

\bibitem{Berger:2019wnu}
N.~Berger et~al., {\it {Simplified Template Cross Sections - Stage 1.1}},
  \href{http://arxiv.org/abs/1906.02754}{{\tt arXiv:1906.02754}}.

\bibitem{ATLAS-CONF-2020-026}
{\bf ATLAS} Collaboration, {\it {Measurement of the properties of Higgs boson
  production at $\sqrt{s}$=13 TeV in the $H\to \gamma\gamma$ channel using 139
  $fb^{-1}$ of $pp$ collision data with the ATLAS experiment}},  tech. rep.,
  CERN, Geneva, 2020.
\newblock All figures including auxiliary figures are available at
  https://atlas.web.cern.ch/Atlas/GROUPS/PHYSICS/CONFNOTES/ATLAS-CONF-2020-026.

\bibitem{ATLAS:2020rej}
{\bf ATLAS} Collaboration, G.~Aad et~al., {\it {Higgs boson production
  cross-section measurements and their EFT interpretation in the $4\ell $ decay
  channel at $\sqrt{s}=$13 TeV with the ATLAS detector}},  {\em Eur. Phys. J.
  C} {\bf 80} (2020), no.~10 957, [\href{http://arxiv.org/abs/2004.03447}{{\tt
  arXiv:2004.03447}}]. [Erratum: Eur.Phys.J.C 81, 29 (2021), Erratum:
  Eur.Phys.J.C 81, 398 (2021)].

\bibitem{ATLAS-CONF-2021-014}
{\bf ATLAS} Collaboration, {\it {Measurements of gluon fusion and
  vector-boson-fusion production of the Higgs boson in $H\rightarrow W W^*
  \rightarrow e\nu \mu\nu$ decays using $pp$ collisions at $\sqrt{s}=13$ TeV
  with the ATLAS detector}},  tech. rep., CERN, Geneva, 2021.
\newblock All figures including auxiliary figures are available at
  https://atlas.web.cern.ch/Atlas/GROUPS/PHYSICS/CONFNOTES/ATLAS-CONF-2021-014.

\bibitem{ATLAS-CONF-2021-044}
{\bf ATLAS} Collaboration, {\it {Measurements of Higgs boson production
  cross-sections in the $H\to\tau^{+}\tau^{-}$ decay channel in $pp$ collisions
  at $\sqrt{s}=13\,\text{TeV}$ with the ATLAS detector}},  tech. rep., CERN,
  Geneva, 2021.
\newblock All figures including auxiliary figures are available at
  https://atlas.web.cern.ch/Atlas/GROUPS/PHYSICS/CONFNOTES/ATLAS-CONF-2021-044.

\bibitem{ATLAS:2020fcp}
{\bf ATLAS} Collaboration, G.~Aad et~al., {\it {Measurements of $WH$ and $ZH$
  production in the $H \rightarrow b\bar{b}$ decay channel in $pp$ collisions
  at 13 TeV with the ATLAS detector}},  {\em Eur. Phys. J. C} {\bf 81} (2021),
  no.~2 178, [\href{http://arxiv.org/abs/2007.02873}{{\tt arXiv:2007.02873}}].

\bibitem{ATLAS:2020jwz}
{\bf ATLAS} Collaboration, G.~Aad et~al., {\it {Measurement of the associated
  production of a Higgs boson decaying into $b$-quarks with a vector boson at
  high transverse momentum in $pp$ collisions at $\sqrt{s} = 13$ TeV with the
  ATLAS detector}},  {\em Phys. Lett. B} {\bf 816} (2021) 136204,
  [\href{http://arxiv.org/abs/2008.02508}{{\tt arXiv:2008.02508}}].

\bibitem{ATLAS-CONF-2021-051}
{\bf ATLAS} Collaboration, {\it {Combination of measurements of Higgs boson
  production in association with a $W$ or $Z$ boson in the $ b\bar{b}$ decay
  channel with the ATLAS experiment at $\sqrt{s}=13$ TeV}},  tech. rep., CERN,
  Geneva, 2021.
\newblock All figures including auxiliary figures are available at
  https://atlas.web.cern.ch/Atlas/GROUPS/PHYSICS/CONFNOTES/ATLAS-CONF-2021-051.

\bibitem{ATLAS:2020bhl}
{\bf ATLAS} Collaboration, G.~Aad et~al., {\it {Measurements of Higgs bosons
  decaying to bottom quarks from vector boson fusion production with the ATLAS
  experiment at $\sqrt{s}=13\,\text {TeV}$}},  {\em Eur. Phys. J. C} {\bf 81}
  (2021), no.~6 537, [\href{http://arxiv.org/abs/2011.08280}{{\tt
  arXiv:2011.08280}}].

\bibitem{ATLAS:2021qou}
{\bf ATLAS} Collaboration, G.~Aad et~al., {\it {Measurement of Higgs boson
  decay into $b$-quarks in associated production with a top-quark pair in $pp$
  collisions at $\sqrt{s}=13$ TeV with the ATLAS detector}},  {\em JHEP} {\bf
  06} (2022) 097, [\href{http://arxiv.org/abs/2111.06712}{{\tt
  arXiv:2111.06712}}].

\bibitem{ATLAS:2019rob}
{\bf ATLAS} Collaboration, M.~Aaboud et~al., {\it {Measurement of fiducial and
  differential $W^+W^-$ production cross-sections at $\sqrt{s}=13$ TeV with the
  ATLAS detector}},  {\em Eur. Phys. J. C} {\bf 79} (2019), no.~10 884,
  [\href{http://arxiv.org/abs/1905.04242}{{\tt arXiv:1905.04242}}].

\bibitem{ATLAS:2019bsc}
{\bf ATLAS} Collaboration, M.~Aaboud et~al., {\it {Measurement of $W^{\pm}Z$
  production cross sections and gauge boson polarisation in $pp$ collisions at
  $\sqrt{s} = 13$ TeV with the ATLAS detector}},  {\em Eur. Phys. J. C} {\bf
  79} (2019), no.~6 535, [\href{http://arxiv.org/abs/1902.05759}{{\tt
  arXiv:1902.05759}}].

\bibitem{ATLAS:2021kog}
{\bf ATLAS} Collaboration, G.~Aad et~al., {\it {Measurements of differential
  cross-sections in four-lepton events in 13 TeV proton-proton collisions with
  the ATLAS detector}},  {\em JHEP} {\bf 07} (2021) 005,
  [\href{http://arxiv.org/abs/2103.01918}{{\tt arXiv:2103.01918}}].

\bibitem{ATLAS:2020nzk}
{\bf ATLAS} Collaboration, G.~Aad et~al., {\it {Differential cross-section
  measurements for the electroweak production of dijets in association with a
  $Z$ boson in proton\textendash{}proton collisions at ATLAS}},  {\em Eur.
  Phys. J. C} {\bf 81} (2021), no.~2 163,
  [\href{http://arxiv.org/abs/2006.15458}{{\tt arXiv:2006.15458}}].

\bibitem{ALEPH:2005ab}
{\bf ALEPH, DELPHI, L3, OPAL, SLD, LEP Electroweak Working Group, SLD
  Electroweak Group, SLD Heavy Flavour Group} Collaboration, S.~Schael et~al.,
  {\it {Precision electroweak measurements on the $Z$ resonance}},  {\em Phys.
  Rept.} {\bf 427} (2006) 257--454,
  [\href{http://arxiv.org/abs/hep-ex/0509008}{{\tt hep-ex/0509008}}].

\bibitem{LHCHiggsCrossSectionWorkingGroup:2016ypw}
{\bf LHC Higgs Cross Section Working Group} Collaboration, D.~de~Florian
  et~al., {\it {Handbook of LHC Higgs Cross Sections: 4. Deciphering the Nature
  of the Higgs Sector}},  \href{http://arxiv.org/abs/1610.07922}{{\tt
  arXiv:1610.07922}}.

\bibitem{Sherpa:2019gpd}
{\bf Sherpa} Collaboration, E.~Bothmann et~al., {\it {Event Generation with
  Sherpa 2.2}},  {\em SciPost Phys.} {\bf 7} (2019), no.~3 034,
  [\href{http://arxiv.org/abs/1905.09127}{{\tt arXiv:1905.09127}}].

\bibitem{Bahr:2008pv}
M.~B{\"a}hr et~al., {\it {Herwig++ Physics and Manual}},  {\em Eur. Phys. J.}
  {\bf C58} (2008) 639--707, [\href{http://arxiv.org/abs/0803.0883}{{\tt
  arXiv:0803.0883}}].

\bibitem{Baglio:2011juf}
J.~Baglio et~al., {\it {VBFNLO: A Parton Level Monte Carlo for Processes with
  Electroweak Bosons -- Manual for Version 2.7.0}},
  \href{http://arxiv.org/abs/1107.4038}{{\tt arXiv:1107.4038}}.

\bibitem{Sjostrand:2014zea}
T.~Sj\"ostrand, S.~Ask, J.~R. Christiansen, R.~Corke, N.~Desai, P.~Ilten,
  S.~Mrenna, S.~Prestel, C.~O. Rasmussen, and P.~Z. Skands, {\it {An
  introduction to PYTHIA 8.2}},  {\em Comput. Phys. Commun.} {\bf 191} (2015)
  159--177, [\href{http://arxiv.org/abs/1410.3012}{{\tt arXiv:1410.3012}}].

\bibitem{Alwall:2011uj}
J.~Alwall, M.~Herquet, F.~Maltoni, O.~Mattelaer, and T.~Stelzer, {\it {MadGraph
  5 : Going Beyond}},  {\em JHEP} {\bf 06} (2011) 128,
  [\href{http://arxiv.org/abs/1106.0522}{{\tt arXiv:1106.0522}}].

\bibitem{Brivio:2017btx}
I.~Brivio, Y.~Jiang, and M.~Trott, {\it {The SMEFTsim package, theory and
  tools}},  {\em JHEP} {\bf 12} (2017) 070,
  [\href{http://arxiv.org/abs/1709.06492}{{\tt arXiv:1709.06492}}].

\bibitem{Brivio:2020onw}
I.~Brivio, {\it {SMEFTsim 3.0 \textemdash{} a practical guide}},  {\em JHEP}
  {\bf 04} (2021) 073, [\href{http://arxiv.org/abs/2012.11343}{{\tt
  arXiv:2012.11343}}].

\bibitem{Dawson:2018liq}
S.~Dawson and P.~P. Giardino, {\it {Electroweak corrections to Higgs boson
  decays to $\gamma\gamma$ and $W^+W^-$ in standard model EFT}},  {\em Phys.
  Rev. D} {\bf 98} (2018), no.~9 095005,
  [\href{http://arxiv.org/abs/1807.11504}{{\tt arXiv:1807.11504}}].

\bibitem{Corbett:2021eux}
T.~Corbett, A.~Helset, A.~Martin, and M.~Trott, {\it {EWPD in the SMEFT to
  dimension eight}},  {\em JHEP} {\bf 06} (2021) 076,
  [\href{http://arxiv.org/abs/2102.02819}{{\tt arXiv:2102.02819}}].

\bibitem{Aoude:2022aro}
R.~Aoude, F.~Maltoni, O.~Mattelaer, C.~Severi, and E.~Vryonidou, {\it
  {Renormalisation group effects on SMEFT interpretations of LHC data}},
  \href{http://arxiv.org/abs/2212.05067}{{\tt arXiv:2212.05067}}.

\bibitem{deBlas:2022ofj}
J.~de~Blas, Y.~Du, C.~Grojean, J.~Gu, V.~Miralles, M.~E. Peskin, J.~Tian,
  M.~Vos, and E.~Vryonidou, {\it {Global SMEFT Fits at Future Colliders}},  in
  {\em {2022 Snowmass Summer Study}}, 6, 2022.
\newblock \href{http://arxiv.org/abs/2206.08326}{{\tt arXiv:2206.08326}}.

\bibitem{deBlas:2019rxi}
J.~de~Blas et~al., {\it {Higgs Boson Studies at Future Particle Colliders}},
  {\em JHEP} {\bf 01} (2020) 139, [\href{http://arxiv.org/abs/1905.03764}{{\tt
  arXiv:1905.03764}}].

\bibitem{Iranipour:2022iak}
S.~Iranipour and M.~Ubiali, {\it {A new generation of simultaneous fits to LHC
  data using deep learning}},  {\em JHEP} {\bf 05} (2022) 032,
  [\href{http://arxiv.org/abs/2201.07240}{{\tt arXiv:2201.07240}}].

\bibitem{Carrazza:2019sec}
S.~Carrazza, C.~Degrande, S.~Iranipour, J.~Rojo, and M.~Ubiali, {\it {Can New
  Physics hide inside the proton?}},  {\em Phys. Rev. Lett.} {\bf 123} (2019),
  no.~13 132001, [\href{http://arxiv.org/abs/1905.05215}{{\tt
  arXiv:1905.05215}}].

\bibitem{Greljo:2021kvv}
A.~Greljo, S.~Iranipour, Z.~Kassabov, M.~Madigan, J.~Moore, J.~Rojo, M.~Ubiali,
  and C.~Voisey, {\it {Parton distributions in the SMEFT from high-energy
  Drell-Yan tails}},  {\em JHEP} {\bf 07} (2021) 122,
  [\href{http://arxiv.org/abs/2104.02723}{{\tt arXiv:2104.02723}}].

\bibitem{Miller:2020hua}
B.~K. Miller, A.~Cole, G.~Louppe, and C.~Weniger, {\it {Simulation-efficient
  marginal posterior estimation with swyft: stop wasting your precious time}},
  \href{http://arxiv.org/abs/2011.13951}{{\tt arXiv:2011.13951}}.

\bibitem{DarkMachinesHighDimensionalSamplingGroup:2021wkt}
{\bf DarkMachines High Dimensional Sampling Group} Collaboration, C.~Bal\'azs
  et~al., {\it {A comparison of optimisation algorithms for high-dimensional
  particle and astrophysics applications}},  {\em JHEP} {\bf 05} (2021) 108,
  [\href{http://arxiv.org/abs/2101.04525}{{\tt arXiv:2101.04525}}].

\bibitem{Carrazza:2021gpx}
S.~Carrazza, J.~Cruz-Martinez, M.~Rossi, and M.~Zaro, {\it {MadFlow: automating
  Monte Carlo simulation on GPU for particle physics processes}},  {\em Eur.
  Phys. J. C} {\bf 81} (2021), no.~7 656,
  [\href{http://arxiv.org/abs/2106.10279}{{\tt arXiv:2106.10279}}].

\bibitem{Carrazza:2020qwu}
S.~Carrazza, J.~M. Cruz-Martinez, and M.~Rossi, {\it {PDFFlow: Parton
  distribution functions on GPU}},  {\em Comput. Phys. Commun.} {\bf 264}
  (2021) 107995, [\href{http://arxiv.org/abs/2009.06635}{{\tt
  arXiv:2009.06635}}].

\end{thebibliography}\endgroup
